\documentclass[preprint,12pt]{elsarticle}

\usepackage{amssymb}
\usepackage{xspace}
\usepackage{url}
\usepackage{tabularx}
\usepackage{multirow}
\usepackage{color}

\usepackage{xcolor}
\usepackage{array}
\usepackage{arydshln}

\usepackage{mathtools}

\usepackage{multirow}

\newcolumntype{L}[1]{>{\raggedright\let\newline\\\arraybackslash\hspace{0pt}}m{#1}}
\newcolumntype{C}[1]{>{\centering\let\newline\\\arraybackslash\hspace{0pt}}m{#1}}
\newcolumntype{R}[1]{>{\raggedleft\let\newline\\\arraybackslash\hspace{0pt}}m{#1}}

\definecolor{darkred}{rgb}{0.44,0,0}
\definecolor{darkgreen}{rgb}{0,0.44,0}
\definecolor{darkblue}{rgb}{0,0,0.44}
\definecolor{grey}{rgb}{0.5,0.5,0.5}

\journal{Parallel Computing}

\begin{document}

\newcommand{\gemm}{{\sc gemm}\xspace}

\begin{frontmatter}



%\title{Performance and Energy Optimization of Matrix \\ Multiplication
%       on Asymmetric big.LITTLE Processors}
\title{Architecture-Aware Configuration and Scheduling of Matrix Multiplication on \\
       Asymmetric Multicore Processors}
       
\author[uji]{Sandra~Catal\'an}
\ead{catalans@uji.es}
\author[ucm]{Francisco~D.~Igual}
\ead{figual@ucm.es}
\author[uji]{Rafael~Mayo}
\ead{mayo@uji.es}
\author[uji]{Rafael~Rodr\'{\i}guez-S\'anchez}
\ead{rarodrig@uji.es}
\author[uji]{Enrique~S.~Quintana-Ort\'{\i}}
\ead{quintana@uji.es}

\address[uji]{Depto. de Ingenier\'{\i}a y Ciencia de Computadores,
             Universidad Jaume I, Castell\'on, Spain.}
\address[ucm]{Depto. de Arquitectura de Computadores y Autom\'atica, Universidad Complutense de Madrid, Spain.}       


\author{}

\address{}

\begin{abstract}
Asymmetric multicore processors (AMPs) have recently emerged as an appealing technology for severely energy-constrained
environments, especially in mobile appliances where heterogeneity in applications is 
mainstream. In addition, given the growing interest for low-power 
high performance computing, this type of architectures is also being investigated 
as a means to improve the throughput-per-Watt of
complex scientific applications. 

In this paper, we design and embed several architecture-aware optimizations into a multi-threaded 
general matrix multiplication (\gemm), a key operation of the BLAS, in order to obtain a high
performance implementation for ARM big.LITTLE AMPs.
Our solution is based on the reference implementation of \gemm in the BLIS library, 
and integrates a cache-aware configuration as well as asymmetric--static and dynamic scheduling
strategies that carefully tune  and distribute the operation's micro-kernels among the big and LITTLE 
cores of the target processor.
The experimental results on a Samsung Exynos 5422, a system-on-chip with ARM Cortex-A15 and Cortex-A7 clusters
that implements the big.LITTLE model, expose that our cache-aware 
versions of \gemm with asymmetric scheduling attain important gains in performance 
with respect to its architecture-oblivious counterparts
while exploiting all the resources of the AMP to deliver considerable energy efficiency.

\end{abstract}


\end{frontmatter}


\section{Introduction}

The decay of Dennard scaling~\cite{Den74} during the past decade
marked the end of the ``GHz race'' and the shift towards multicore designs 
due to their more favorable performance-power ratio.
In addition, the doubling of transistors on chip with each new semiconductor generation,
dictated by Moore's law~\cite{Moo65}, has only exacerbated the 
{\em power wall} problem~\cite{Dur13,Lec13,Luc14}, leading to
the arise of ``dark silicon''~\cite{Esm11} 
and the deployment of heterogeneous facilities for high performance computing~\cite{top500,green500}.

Asymmetric multicore processors (AMPs) are a particular class 
of heterogeneous architectures equipped 
with cores that share the same instruction set architecture\footnote{According
to this definition, servers 
equipped with one (or more) general-purpose multicore processor(s) 
and a PCIe-attached graphics accelerator, or systems-on-chip like the NVIDIA
Tegra TK1, are excluded from this category.} but
differ in micro-architecture, and thus in complexity, performance, and power consumption. 
AMPs have received considerable attention in the last years
as a means to improve the performance-power ratio of
computing systems~\cite{Kum04,Hil08,Mor06,Win10} partly because, in theory, 
they can deliver much higher performance for the same
power budget, mainly by exploiting the presence of serial and parallel phases within
applications% and different workload characterization
~\cite{Mor06}.

The {\em general matrix multiplication} (\gemm) is a crucial operation for the optimization of the Level-3 {\em Basic Linear Algebra Subprograms} (BLAS)~\cite{blas3}, 
as portable and highly tuned versions of the remaining Level-3 kernels are in general built on top of \gemm~\cite{Kagstrom:1998}.
In turn, the contents of BLAS conform a pivotal cornerstone upon which 
many sophisticated libraries to tackle complex scientific and engineering applications rely~\cite{AsaBCGHKPPSWY06}. 
The importance of BLAS in general, and \gemm in particular, is illustrated by the 
prolonged efforts spent over the past decades to produce carefully tuned %implementations in the form of 
commercial libraries for almost any current architecture
(e.g., Intel's MKL~\cite{MKL}, AMD's ACML~\cite{ACML}, IBM's ESSL~\cite{ESSL}, NVIDIA's CUBLAS~\cite{cublas}, etc.)
as well as the number of high quality open source solutions  (e.g., GotoBLAS~\cite{Goto:2008:AHP,Goto:2008:HPI},
OpenBLAS~\cite{OpenBLAS}, BLIS~\cite{BLIS1}, and ATLAS~\cite{atlas}).

%ARM's asymmetric big.LITTLE technology, employing as a case of study the compute-intensive general matrix multiplication (\gemm):
%$C \mathrel{+}= A \cdot B$, where the sizes of $A$, $B$, $C$ are
%respectively $m \times k$, $k \times n$, $m \times n$.

In this paper we propose efficient multi-threaded implementations of \gemm on 
an ARM big.LITTLE AMP consisting of a cluster composed of a few fast (big) cores 
and a complementary cluster with several slow (LITTLE) cores, 
shared main memory, and private L1/L2 caches per core/cluster, respectively.
%In an abstract point of view,
Our approach leverages the multi-threaded implementation of \gemm in the BLIS library,
which decomposes the operation into a collection of nested loops around a {\em micro-kernel}.
In this reference code, we modify the loop stride configuration and scheduling 
to distribute the micro-kernels comprised by certain loops among the big/LITTLE clusters and cores while taking into account
the processor's computational power and cache organization.
In more detail, this work makes the following specific contributions:
\begin{itemize}
\item Our optimized implementations modify the control tree structure that governs the multi-threaded parallelization of BLIS \gemm 
      in order to accommodate cache-aware configurations of the loop strides for each type
      of core architecture that match the organization of its cache hierarchy.
\item We integrate two alternative scheduling strategies, asymmetric--static and dynamic, 
      to produce a 1-D partitioning of (the iteration space for) one of the outer loops of BLIS \gemm
      between the two clusters that yields a balanced distribution of the micro-kernels. Furthermore, we apply an orthogonal symmetric--static schedule 
      to map the workload of one of the inner loops
      across the cores of the same cluster. 
\item We demonstrate the practical benefits of the cache-aware configurations and asymmetry-aware 
      scheduling strategies on the Exynos 5422, a system-on-chip (SoC) consisting of
      an ARM Cortex-A15 quad core (big) cluster and an ARM Cortex-A7 quad core (LITTLE) cluster. 
      Our experimental results show that the performance attained by the optimized \gemm on this platform 
      is well beyond that of an architecture-oblivious multi-threaded implementation and close to that of an ideal scenario. 
\item We include an analysis of the energy efficiency of the asymmetric architecture when running our optimized \gemm,
      using the GFLOPS/W (billions of floating-point arithmetic operations, or flops, per second and Watt) metric,
      which assesses the energy cost of flops.
\end{itemize}
To conclude, we emphasize that the scheduling approaches proposed in this paper
are general and, in combination with the BLIS implementation of \gemm,  
can be ported with little effort to present and future instances of the ARM big.LITTLE architecture
as well as to any other asymmetric design in general (e.g. the Intel QuickIA prototype~\cite{IntelQuickIA}).
Furthermore, we are confident that the principles underlying our scheduling decisions 
carry over to all Level-3 BLAS operations.

The rest of the paper is structured as follows. 
In Section~\ref{sec:related}, we compare our approach to 
optimize {\sc gemm} on AMPs with state-of-the-art works on similar architectures.
In Section~\ref{sec:blis}, we describe the mechanisms that underlie the original multi-threaded implementation 
of {\sc gemm} in the BLIS framework, and evaluate its performance and optimal cache parameter configuration for the Cortex-A15 
and Cortex-A7 clusters.
In Section~\ref{sec:performance}, we investigate the effect of using standard, architecture-oblivious multi-threaded
BLAS implementations on AMPs, and its negative impact on performance and energy efficiency.
In Section~\ref{sec:strategies}, we introduce our strategies to adapt the original BLIS 
multi-threaded implementation to the asymmetric architecture, and report the performance and energy-efficiency
results of the new codes.
Finally, Section~\ref{sec:conclusions} closes the paper with a few concluding remarks and proposals for future work.

\section{Related Work}
\label{sec:related}

Heterogeneous (and asymmetric) architectures are an active research topic,
with a vast design space that needs careful consideration in terms of power, performance, 
programmability, and flexibility~\cite{Chi12}. 
Many of these works can be grouped into 
{\em i)} efforts to experimentally evaluate the
computational performance and/or power-energy efficiency of
AMPs using multi-threaded benchmarks and applications; and 
{\em ii)} contributions related to workload-partitioning
strategies for the execution of \gemm on heterogeneous platforms.
In the first group, Winter et al.~\cite{Win10} 
discuss power management techniques and thread scheduling for AMPs;
and scheduling on AMP architectures is explored in a number of works; see,
among others,~\cite{Hou09,Lak09,Rod13,Win10} and the references therein.

In the second group, mapping \gemm in an heterogeneous cluster is analyzed
in~\cite{Cla11}, while a theoretical study of dynamic scheduling applied
to \gemm in a similar scenario is introduced in~\cite{Bea14}. 

Compared with previous work, our investigation aims to deliver an implementation of \gemm,
based on an open source BLAS library (BLIS),
that is highly optimized for asymmetric ARM big.LITTLE architectures.
All previous efforts to implement and evaluate \gemm on AMPs employ simple
codes, at best tuned via very basic tiling techniques, and therefore yield suboptimal codes. 
The research with heterogeneous clusters targets a more general and complex problem, and in practice can 
hardly be expected to produce an optimal solution for AMPs.

\section{Multi-Threaded Portable Implementation of BLIS \gemm}
\label{sec:blis}

Modern high-performance implementations of \gemm for general-purpose architectures
%, including BLIS and OpenBLAS,
follow the design pioneered by GotoBLAS~\cite{Goto:2008:AHP}.
BLIS in particular implements the \gemm 
$C \mathrel{+}= A \cdot B$, where the sizes of $A$, $B$, $C$ are 
respectively $m \times k$, $k \times n$, $m \times n$,
as three nested loops around a {\em macro-kernel} plus two packing routines
(see Loops~1--3 in Figure~\ref{fig:gotoblas_gemm}).
The macro-kernel is then implemented in terms of two additional loops around a {\em
micro-kernel} (Loops~4 and~5 in Figure~\ref{fig:gotoblas_gemm}). 
%Finally, the micro-kernel is a loop around a rank-1 (i.e.,
%outer product) update (Loop~6 in Figure~\ref{fig:gotoblas_gemm}). 
%In the figure, $C_c \equiv C(i_c:i_c+m_c-1,j_c:j_c+n_c-1)$ 
%is just a notation artifact, 
%introduced to ease the presentation of the algorithm, 
%while $A_c,B_c$ correspond to actual buffers that are involved in data copies.
In BLIS, the micro-kernel is typically encoded as a loop around a 
rank--1 (i.e., outer product) update using assembly or
with vector intrinsics, while the remaining five loops and packing routines are implemented in C;
see~\cite{BLIS1} for further details.

\begin{figure}[!htb]
\begin{center}
\begin{tabular}{c}
\begin{minipage}[c]{0.75\textwidth}
\includegraphics[width=\textwidth]{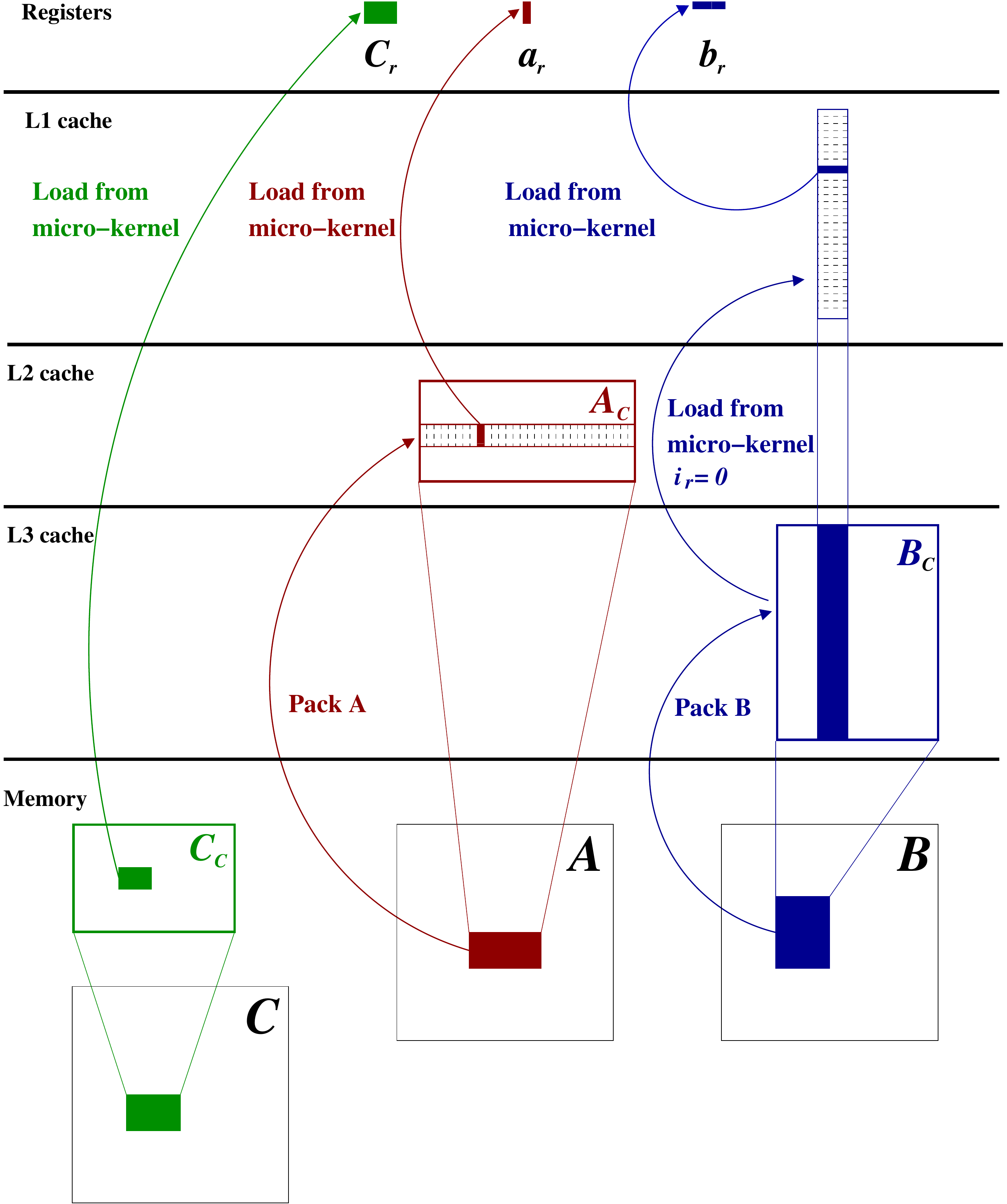}
\end{minipage}
\end{tabular}
\end{center}
\caption{Data movement involved in the BLIS implementation of \gemm.}
\label{fig:movements_gemm}
\end{figure}

Figure~\ref{fig:movements_gemm} illustrates how the loop ordering, together with the packing
routines and an appropriate choice of the  BLIS cache configuration parameters orchestrate
a regular pattern of data transfers through the levels of the memory
hierarchy. In practice, the cache parameters $n_c$, $k_c$, $m_c$, $n_r$ and $m_r$ (which dictate the strides 
of the five outermost loops)
are adjusted taking into account the latency of the floating-point units (FPUs),
number of vector registers, and size/associativity degree of the cache levels.
The goal is that a $k_c \times n_r$ micro-panel of $B_c$, say $B_r$, 
and the $m_c \times k_c$ macro-panel $A_c$ are streamed into the  FPUs from the L1 and L2 caches, respectively;
while the $k_c \times n_c$ macro-panel $B_c$ resides in the L3 cache (if present).
%Note that $C_c$ is just a notation artifact, introduced to ease the presentation of
%the algorithm, while $A_c$ and $B_c$ correspond to actual buffers, with contiguous
%storage in memory, that are involved in copy and packing operations.
By appropriately choosing the configuration parameters,
these transfers are fully amortized with enough computation from within
the micro-kernel; see~\cite{BLIS4}.

% The
%goal is that $A_c$ and a narrow column panel of $B_c$, say $B_r$, are feed into
%the floating-point units from the L2 and L1 caches, respectively, and these
%transfers are fully amortized with enough computation from within
%the micro-kernel; see~\cite{BLIS4}.

\begin{figure}[t]
\centering
%\begin{tabular}{c}
%\begin{minipage}[c]{0.9\textwidth}
%\includegraphics[width=0.9\textwidth]{Figures/gotoblas_gemm.pdf}
%\end{minipage}\\
%\\
\begin{minipage}[c]{\textwidth}
\footnotesize
\resizebox{\linewidth}{!}{
\begin{tabular}{llll}
Loop 1 &{\bf for} $j_c$ = $0,\ldots,n-1$ {\bf in steps of} $n_c$\\
Loop 2 & \hspace{3ex}  {\bf for} $p_c$ = $0,\ldots,k-1$ {\bf in steps of} $k_c$\\
&\hspace{6ex}           \textcolor{darkblue}{$B(p_c:p_c+k_c-1,j_c:j_c+n_c-1)$} $\rightarrow \textcolor{darkblue}{B_c}$ & & // Pack into $B_c$\\
Loop 3 & \hspace{6ex}           {\bf for} $i_c$ = $0,\ldots,m-1$ {\bf in steps of} $m_c$\\
&\hspace{9ex}                     \textcolor{darkred}{$A(i_c:i_c+m_c-1,p_c:p_c+k_c-1)$} $\rightarrow \textcolor{darkred}{A_c}$ & & // Pack into $A_c$ \\
\cline{2-4} 
Loop 4&\hspace{9ex} {\bf for} $j_r$ = $0,\ldots,n_c-1$ {\bf in steps of} $n_r$  & & // Macro-kernel\\
Loop 5&\hspace{12ex}   {\bf for} $i_r$ = $0,\ldots,m_c-1$ {\bf in steps of} $m_r$\\
\cline{2-3}
%Loop 6&\hspace{15ex}    {\bf for} $p_r$ = $0,\ldots,k_c-1$ {\bf in steps of} $1$ & // Micro-kernel \\
&\hspace{15ex}             \textcolor{darkgreen}{$C_c(i_r:i_r+m_r-1,j_r:j_r+n_r-1)$} & // Micro-kernel \\
%&\hspace{19ex} $:=$ ~\textcolor{darkgreen}{$C_c(i_r:i_r+m_r-1,j_r:j_r+n_r-1)$}\\
&\hspace{19ex} ~$\mathrel{+}=$     ~\textcolor{darkred}{$A_c(i_r:i_r+m_r-1,0:k_c-1)$} \\
%&\hspace{19ex} ~~~$\cdot$\!~~~~\textcolor{darkblue}{$B_c(0:k_c-1,j_r:j_r+n_r-1)$} \\
&\hspace{19ex} ~~~$\cdot$~~~~\textcolor{darkblue}{$B_c(0:k_c-1,j_r:j_r+n_r-1)$} \\
%&\hspace{15ex} {\bf endfor}\\
\cline{2-3}
&\hspace{12ex} {\bf endfor}\\
&\hspace{9ex} {\bf endfor}\\
\cline{2-4} 
&\hspace{6ex} {\bf endfor}\\
&\hspace{3ex} {\bf endfor}\\ 
&{\bf endfor}\\ 
\end{tabular}
}
\end{minipage}
%\end{tabular}
\caption{High performance implementation of \gemm in BLIS. In the code, $C_c \equiv C(i_c:i_c+m_c-1,j_c:j_c+n_c-1)$
is just a notation artifact, introduced to ease the presentation of the algorithm, while $A_c,B_c$ correspond to actual buffers that are involved in data copies.}
\label{fig:gotoblas_gemm}
\end{figure}

\subsection{Multi-threaded \gemm in BLIS}
\label{Sec:Parallelization}

BLIS allows to select, at execution time, which
of the five loops of \gemm are parallelized.
Several loops can be simultaneously executed in parallel in
order to adapt the execution to specific properties of the target architecture.
By default, when one of the loops is parallelized, a static partitioning and
mapping of iteration chunks to threads is performed prior to the execution of the loop.

The multi-threaded version of \gemm integrated in BLIS has been previously analyzed 
for conventional symmetric multicore processors (SMPs)~\cite{BLIS2} and modern many-threaded architectures~\cite{BLIS3}.
%such as the IBM PowerPC A2 (16 cores/64 threads) and the Intel Xeon Phi (60 cores/240 threads).
In both ``types'' of architectures, the parallel implementations exploit
the concurrency available in
the nested five--loop organization of \gemm at one or more levels
(i.e., loops). Furthermore,
the approach takes into account the cache organization of the target platform (e.g.,
the presence of multiple sockets, which cache levels are shared/private, etc.), 
while discarding the parallelization of loops that would incur into race conditions
as well as loops with options that exhibit too-fine granularity.
The insights gained from these analyses~\cite{BLIS2,BLIS3} about the loop(s) to parallelize 
in a multi-threaded implementation of \gemm can be summarized as follows:
\begin{itemize}
\item Loop~5 (indexed by $i_r$). With this option,
      different threads execute independent instances of the micro-kernel,
      while accessing the same micro-panel $B_r$ in the L1 cache. 
      The amount of parallelism in this case,
      $\lceil \frac{m_c}{m_r} \rceil$, 
      is scarce as, for many architectures, the optimal value for $m_c$ is in the order of a few hundreds. 
      %Thus, if parallelized, the cost of moving $B_r$ into the L1 cache may not
      %be amortized over a sufficiently large number of flops.
\item Loop~4 (indexed by $j_r$).
      Different threads operate on independent instances of the micro-kernel, but
      access the same macro-panel $A_c$ in the L2 cache.
      The time spent in this loop amortizes the cost of packing (and, therefore, moving)
      $A_c$ from main memory into the L2 cache.
      The amount of parallelism, $\lceil \frac{n_c}{n_r} \rceil$,
      is in general larger than in the previous case, 
      as $n_c$ is in the order of several hundreds up to 
      a few thousands for many architectures.
\item Loop~3 (indexed by $i_c$).
      Each thread packs a different macro-panel $A_c$ into the L2 cache
      and executes a different instance of the macro-kernel.
      The number of iterations of this loop is not
      constrained by the cache parameters, but instead depends on the problem dimension $m$. 
      When $m$ is less than the product of $m_c$ and the degree of parallelization of the loop, 
      $A_c$ will be smaller than the optimal dimension and performance may suffer.
      When there is a shared L2 cache,
      the size of $A_c$ will have to be reduced by a 
      factor equal to the degree of parallelization of this loop. 
      However, reducing $m_c$ is equivalent to parallelizing the
      first loop around the micro-kernel.
\item Loop~2 (indexed by $p_c$). This is not a good choice
      because multiple threads simultaneously update the same parts of 
      $C$, requiring a mechanism to prevent race conditions. 
\item Loop~1 (indexed by $j_c$).
      From a data-sharing perspective, this option is
      equivalent to extracting the parallelism outside of BLIS.
      This parallelization is reasonable in a 
      multi-socket system where each CPU (socket) has a separate L3 cache. 
\end{itemize}

To sum up, these are general guidelines to decide which loops are theoretically good candidates to be parallelized in order
to fully exploit the cache hierarchy of a target architecture. 
At a glance, the appropriate combination of loops to parallelize strongly depends on which caches are private or shared. 
Usually, Loop~1 is a good candidate 
in a multi-socket platform with on-chip L3 caches; 
Loop~3 should be parallelized when 
each core has its own L2 cache; and
Loops~4 and~5 
are convenient choices if the cores share the L2 cache.

\subsection{Experimental setup}

\begin{figure}[t]
\begin{center}
\includegraphics[width=0.6\columnwidth]{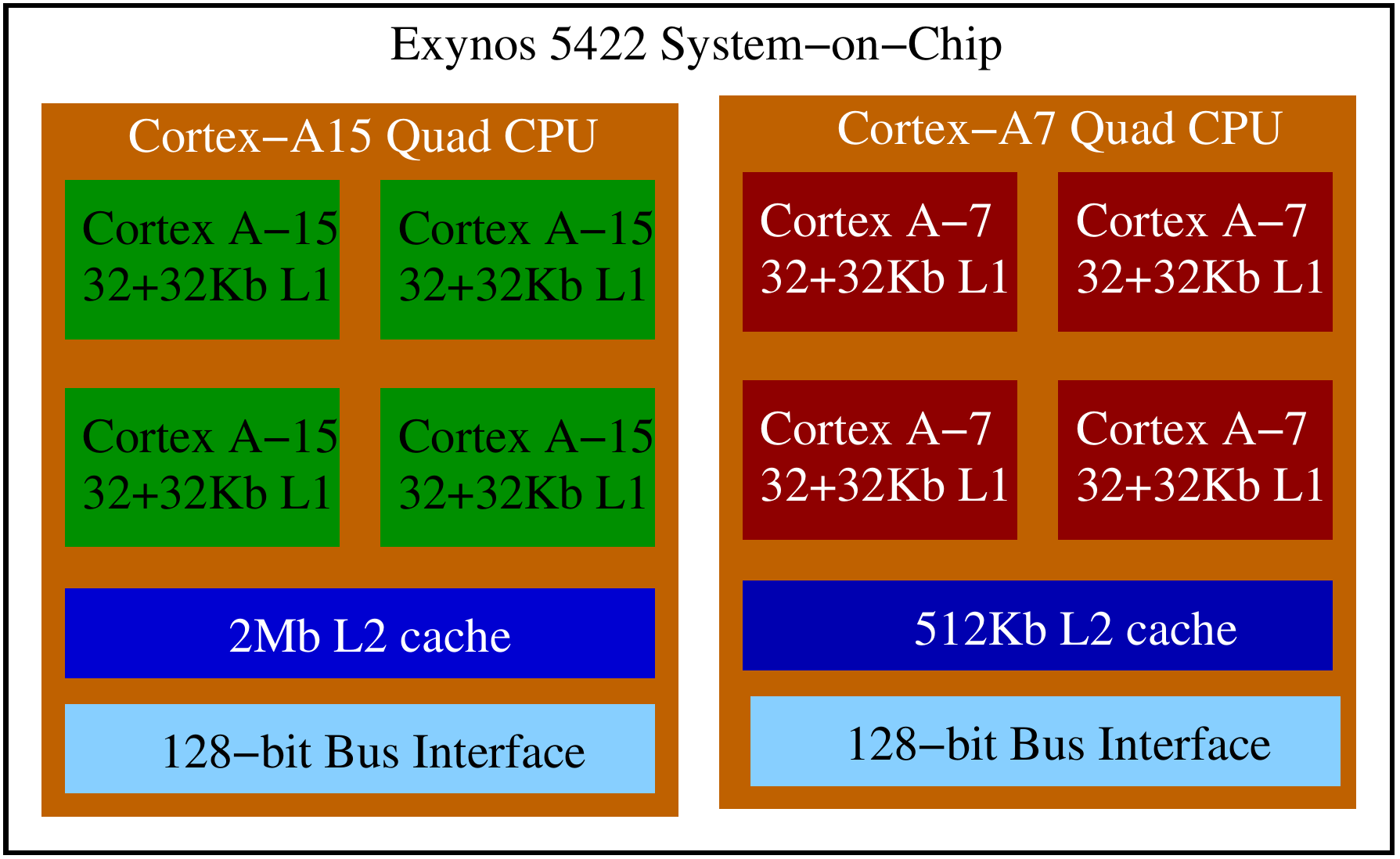}
\end{center}
\caption{\label{fig:exynos} Exynos 5422 block diagram.}
\end{figure}

The ODROID-XU3 board employed in our experiments contains a Samsung Exynos 5422 SoC with an ARM
Cortex-A15 quad-core processing cluster (running at 1.6~GHz in our setup) and a Cortex-A7 quad-core processing 
cluster (running at 1.4~GHz).  Both clusters access a shared DDR3 RAM (2~Gbytes) via 128-bit coherent bus interfaces.
Each ARM core (either Cortex-A15 or Cortex-A7) has a 32+32-Kbyte L1 (instruction+data) cache.
The four ARM Cortex-A15 cores share a 2-Mbyte L2 cache, while the four ARM Cortex-A7 cores
share a smaller 512-Kbyte L2 cache; see Figure~\ref{fig:exynos}.
All our tests hereafter employ {\sc ieee} double-precision arithmetic and square matrices of order $r=m=n=k$. 
We  ensure that the cores 
run at their highest frequency by setting the Linux {\em performance} governor with the appropriate
frequency limits. 
Codes are instrumented with the {\tt pmlib}~\cite{AlonsoICPP12} framework, which 
collects power consumption data corresponding 
to instantaneous power readings from four independent sensors in the board 
(for the Cortex-A7 cores, Cortex-A15 cores, DRAM and GPU), with a sampling rate of 250~ms.

\subsection{Cache optimization for the big and LITTLE cores}

An initial step in order to attain high performance with the implementation of BLIS \gemm is, given a target precision (single or double), 
determine the configuration parameters $n_c$, $k_c$, $m_c$, $n_r$, and $m_r$
for a single ARM core of each type, Cortex-A15 and Cortex-A7, that fit the cache organization.
We next describe our experimental effort
towards this goal. % using {\sc ieee} 754 double-precision arithmetic.
A recent study~\cite{FLAWN74} shows that, in principle, 
this optimization is also possible via analytic derivation.

%\subsection{Optimization of GEMM's cache parameters for the single-core}
%
%{\bf Missing: a 2-D plot with colors to illustrate the effect of $k_c$ and $m_c$, 
%like the one used for the paper on analytical modeling of BLIS?}

\begin{figure}[t]%[h]
\begin{center}
\begin{tabular}{c}
\begin{minipage}[c]{\columnwidth}
\includegraphics[width=0.5\columnwidth]{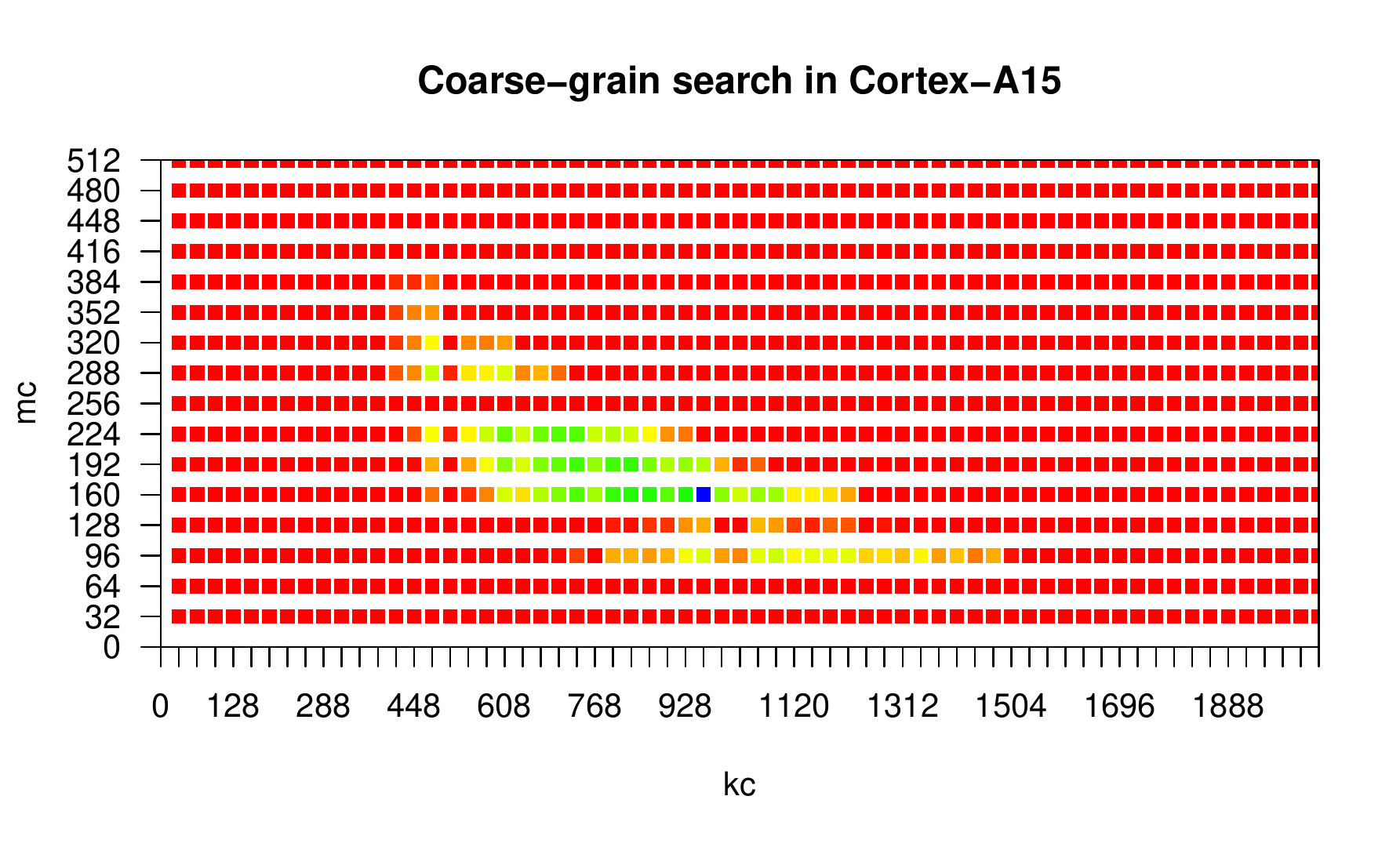}
\includegraphics[width=0.5\columnwidth]{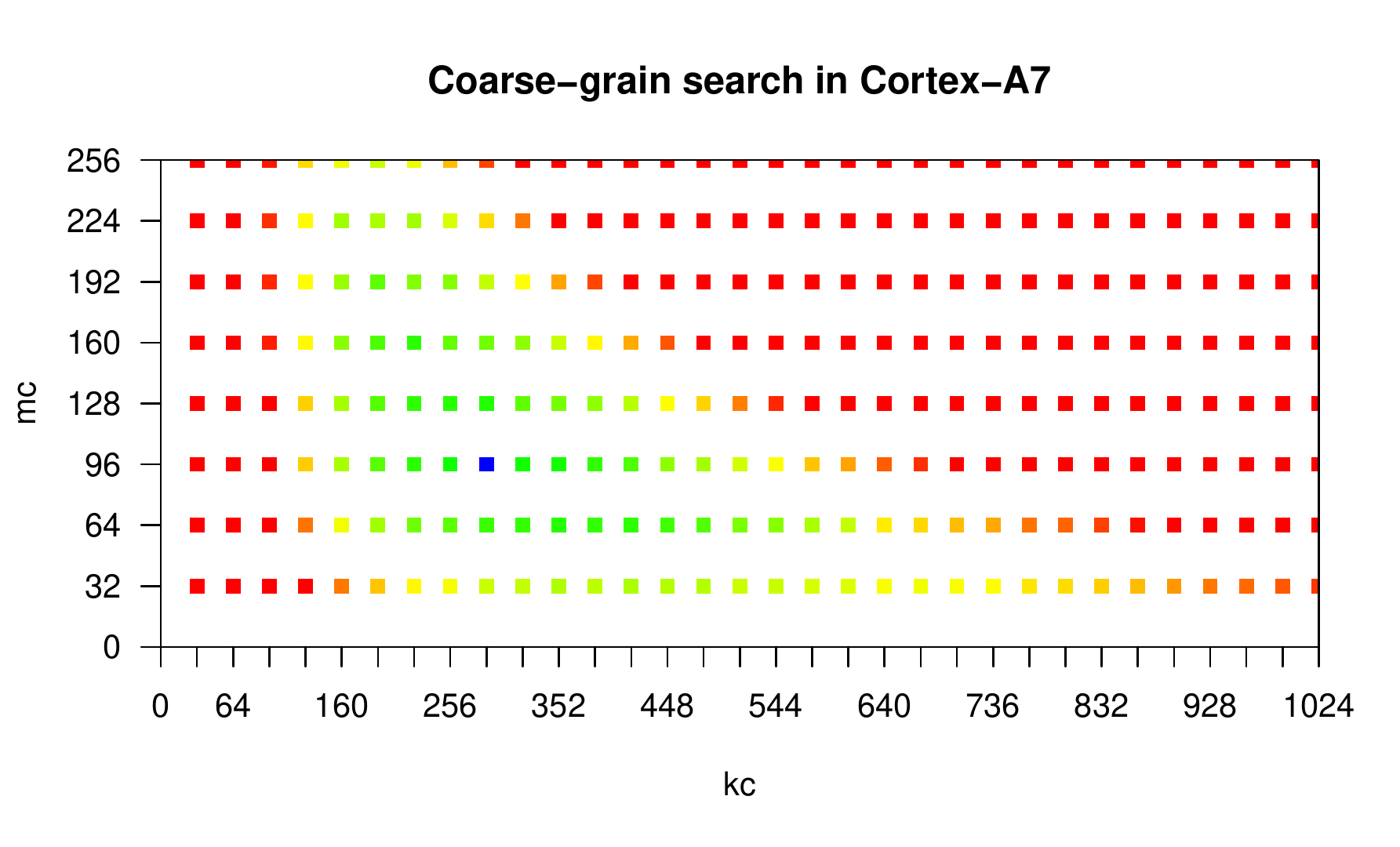}\\
\includegraphics[width=0.5\columnwidth]{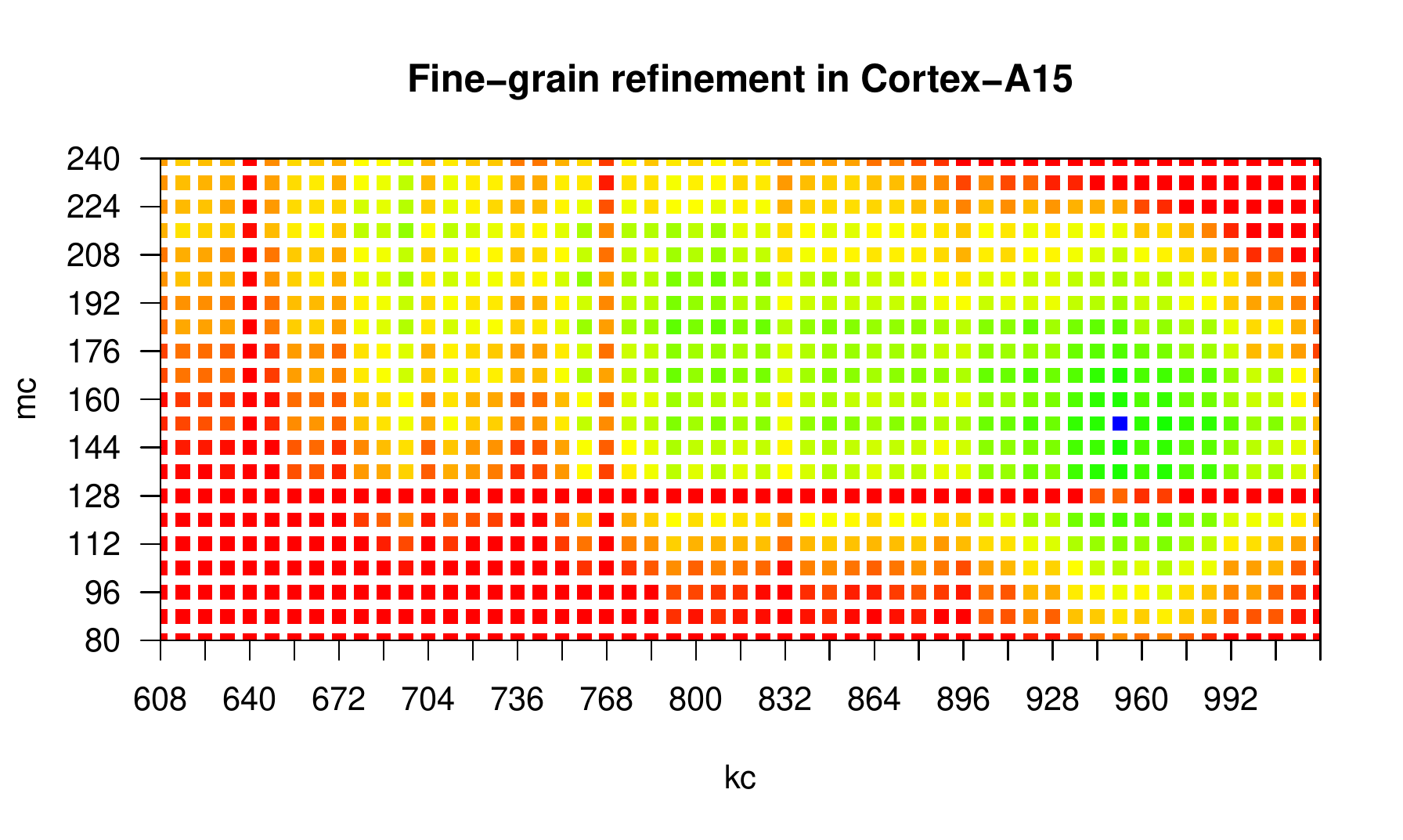}
\includegraphics[width=0.5\columnwidth]{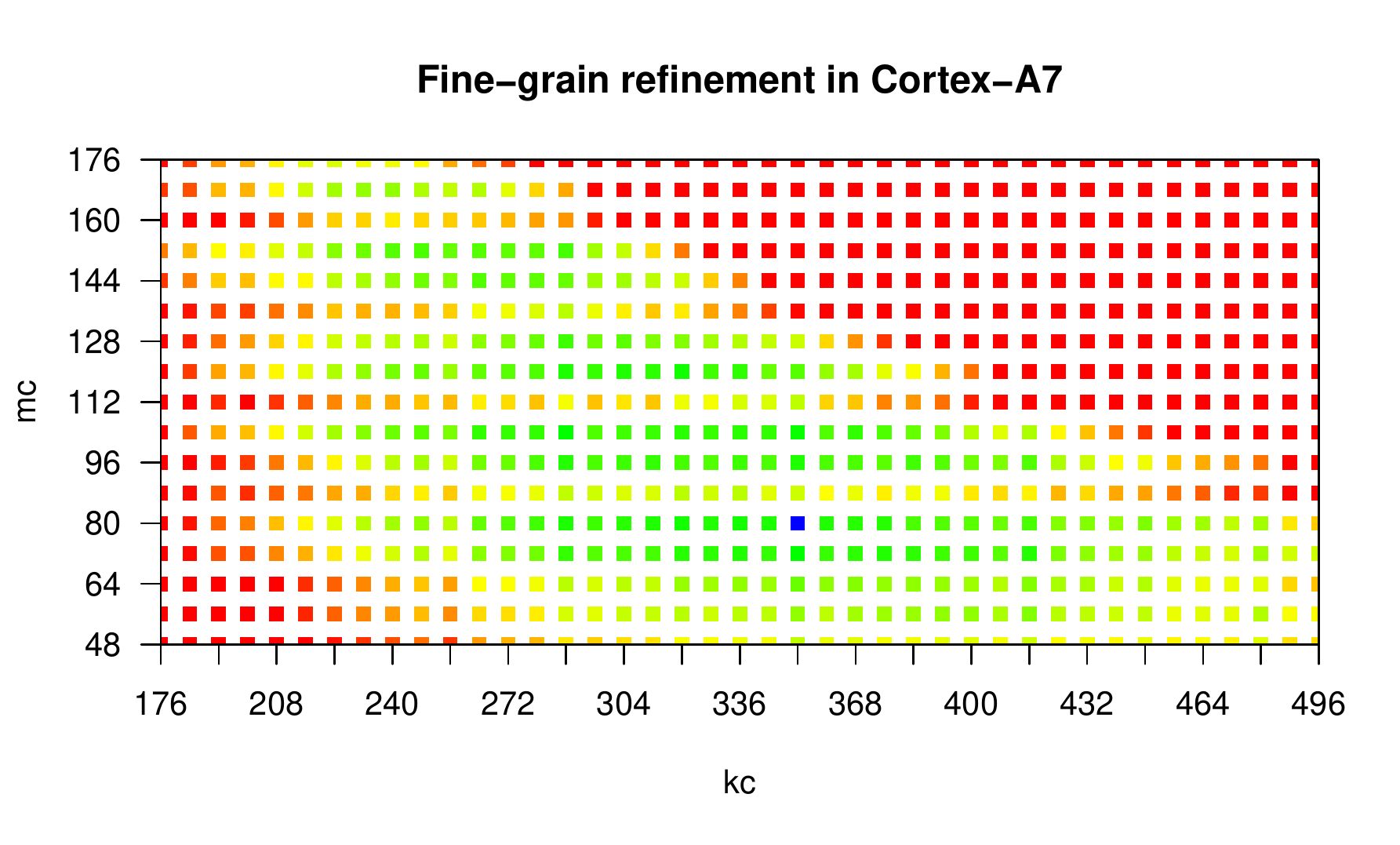}
\end{minipage}
\end{tabular}
\end{center}
\caption{BLIS optimal cache configuration parameters $m_c$ and $k_c$ for the ARM Cortex-A15 (left) and Cortex-A7 (right) in the Samsung Exynos 5422 SoC.
The performance ranges from red (lowest GFLOPS) to green (highest GFLOPS); the optimal ($m_c$, $k_c$) pair is marked
as a blue dot.}
\label{fig:Optimal}
\end{figure}

The first aspect to note is that, in this architecture, $n_c$ plays a minor role and, therefore, can be simply set to
$n_c = 4,096$. This is explained because, in BLIS, $n_c$ is connected to the dimension of 
the L3 cache, which is not present in the Exynos 5422 SoC. 
Furthermore, the micro-kernels for these core architectures
and precision are thoroughly tuned with $m_r = 4$ and $n_r = 4$. 
%These optimal values are used in this work for both the Cortex-A7 and the Cortex-A15 cores.
In consequence, the optimization of \gemm in a single-core scenario 
boils down to determining the optimal values of $m_c$ and $k_c$ for each type of core.
For this purpose, we performed independent empirical searches using a single Cortex-A15 core and 
a single Cortex-A7 core.
In both cases, we initially applied a coarse-grain search to detect potential optimal regions,
and the selected regions were further explored next with a finer granularity 
to detect the optimal configuration parameters. 
The result of this process is illustrated in Figure~\ref{fig:Optimal}, where the top and bottom plots correspond to
the coarse search and the fine-grain refinement respectively.
Performance is measured hereafter in terms of GFLOPS.

The optimal configurations were detected at 
$m_c = 152,~k_c = 952$ for the Cortex-A15 core and
$m_c = 80,~k_c = 352$ for the Cortex-A7 core.
As could be expected, the optimal values for the Cortex-A15 core are larger than those
of  the Cortex-A7 core, since the L2 cache of the former is four times bigger. For both types of cores, the corresponding
dimensions and the associativy-degree of the caches 
allow that the micro-panel $B_r$ ($k_c \times n_r$) fits into the L1 cache 
while the macro-panel $A_c$ ($m_c \times k_c$) resides into the L2 cache.

%\begin{figure}[t]
%\begin{center}
%\begin{tabular}{c}
%\begin{minipage}[c]{\textwidth}
%\includegraphics[width=0.5\textwidth]{./Graph/Cortex-A15}
%\includegraphics[width=0.5\textwidth]{./Graph/Cortex-A15-refinado-kc1024}
%\end{minipage}
%\end{tabular}
%\end{center}
%\caption{Optimal block sizes of the Cortex-A15 cluster.}
%\label{fig:Cortex-a7}
%\end{figure}

\subsection{Multi-threaded BLIS performance on the big and LITTLE clusters}

After determining the optimal configuration parameters for each core cache organization,
we analyze the performance and energy efficiency of a multi-threaded implementation of BLIS \gemm
that operates in a homogeneous (symmetric) system consisting of up to four cores from either the Cortex-A15 cluster or the Cortex-A7 cluster. 
In particular, given the guidelines summarized in Section~\ref{Sec:Parallelization}, and the fact that the L2 cache
is shared among the cores of the same cluster, we adopt a static parallelization of Loop~4 
using 1--4 threads/cores. Similar qualitative conclusions were obtained from a static
parallelization of Loop~5. We note that, although the two types of clusters are evaluated in isolation in this
section, the performance GFLOPS figures will be of interest for the asymmetric-aware versions of {\sc gemm}
that will be presented in Sections~\ref{sec:performance} and \ref{sec:strategies}, as their aggregation can be considered as an {\em ideal scenario}
for the peak performance that can be extracted from the complete asymmetric SoC.

\begin{figure}[t]
\begin{center}
\begin{tabular}{c}
\begin{minipage}[c]{\columnwidth}
\includegraphics[width=0.5\columnwidth]{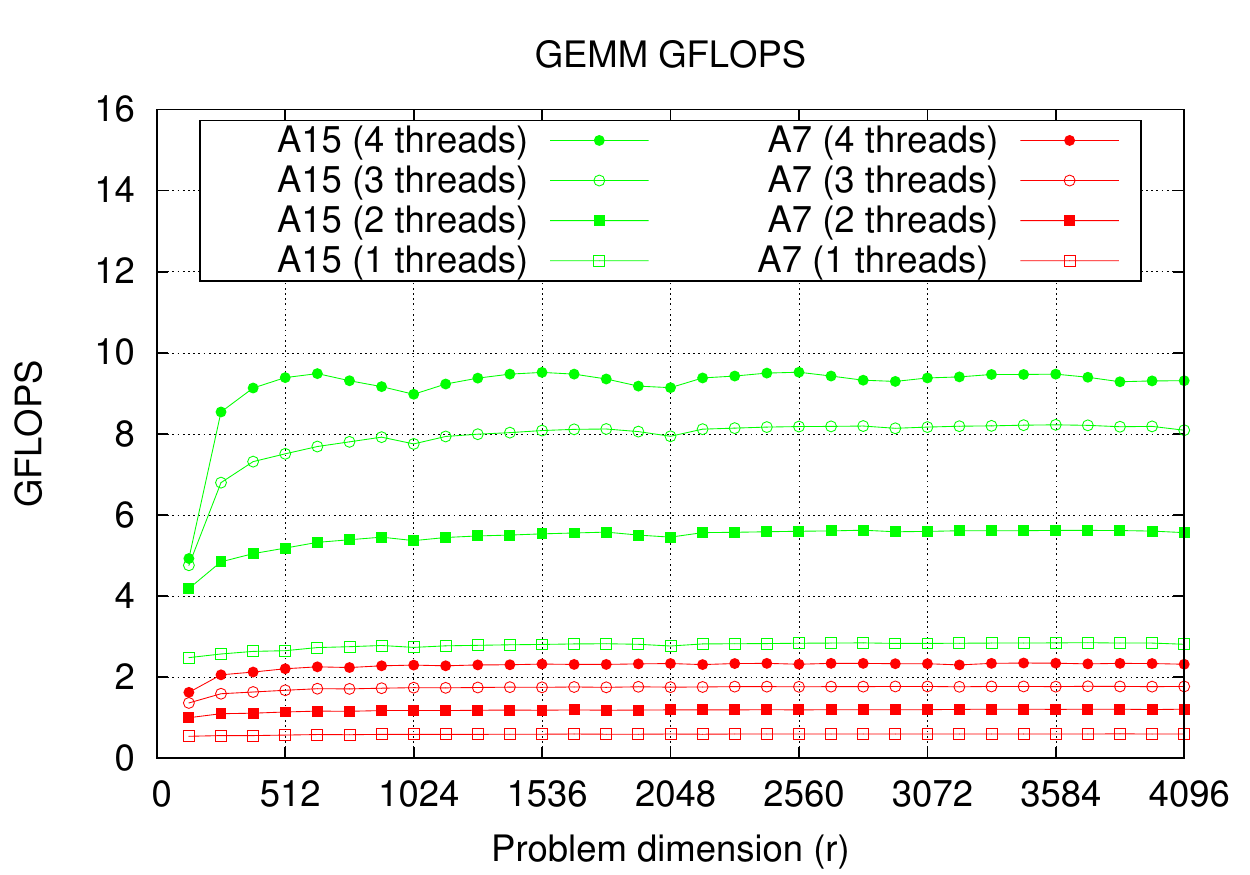}
\includegraphics[width=0.5\columnwidth]{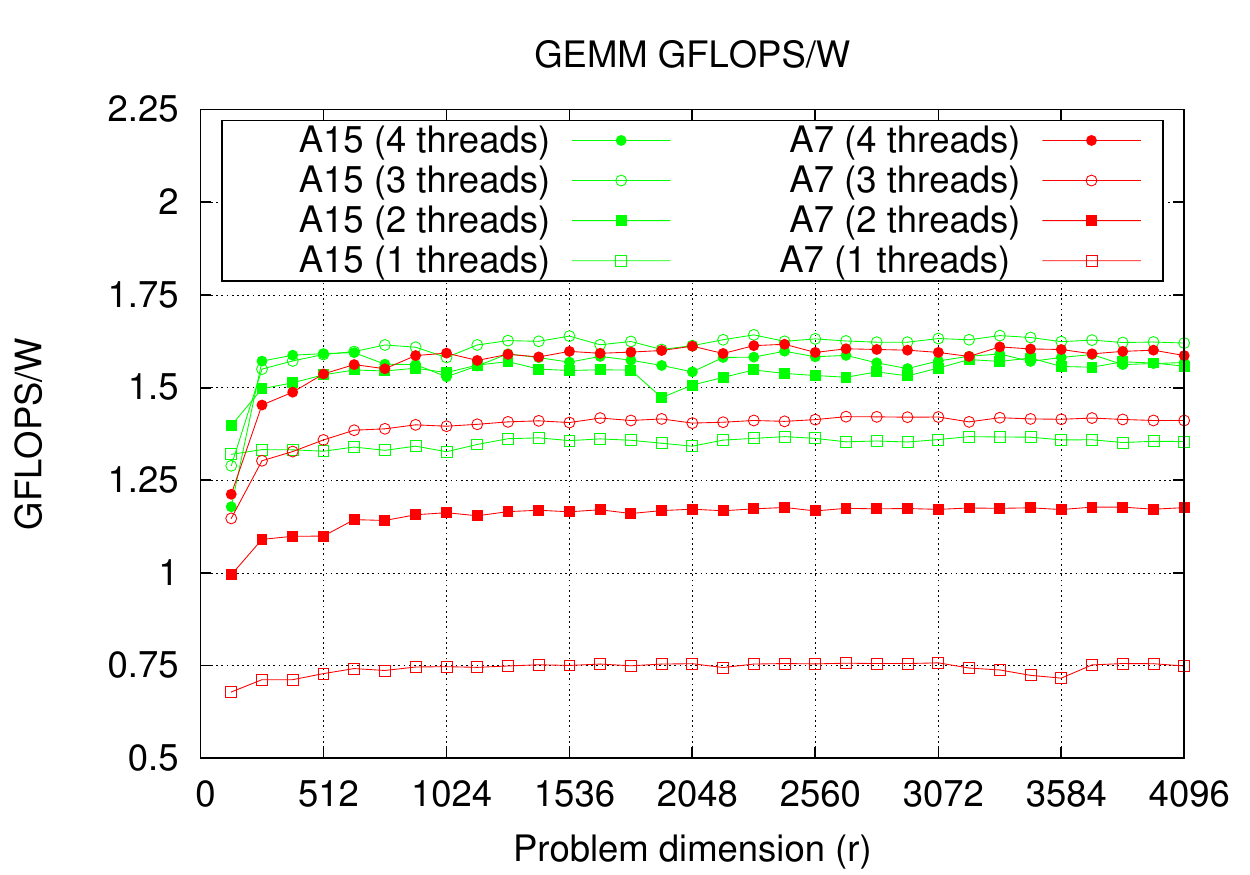}
\end{minipage}
\end{tabular}
\end{center}
\caption{Performance (left) and energy efficiency (right) of the BLIS \gemm using 
exclusively one type of core, for a varying number of threads.}
\label{fig:threads}
\end{figure}

The plots in Figure~\ref{fig:threads} show the performance and energy efficiency 
of the multi-threaded {\sc gemm} implementation in BLIS when using the Cortex-A15 and the Cortex-A7 clusters 
in isolation. 
We note that, when calculating the energy efficiency of one type of cluster, 
the energy consumed by the complementary (idle) cluster is also accounted for, so that we are reporting the energy efficiency of
the complete SoC.

Focusing on performance first, the results expose that the Cortex-A15 cores 
deliver considerable higher performance than their Cortex-A7 counterparts. %which is not surprising. 
Specifically, the former type of cores renders an increase of $2.8$ GFLOPS per added core 
when up to three cores are used, though the utilization of the fourth core yields a smaller increase,  of an 
additional $1.4$ GFLOPS.
%since some computational resources should be dedicated to the operating system.
In conjunction, the four cores of the Cortex-A15 cluster attain a peak performance of $9.6$ GFLOPS. 
For the Cortex-A7 cluster, the peak performance is close to $2.4$ GFLOPS, also attained with four cores. 

Regarding energy efficiency, the Cortex-A15 offers the best results in terms of GFLOPS/W. However, the benefits of
increasing the number of threads are less significant  
when compared with those obtained with the Cortex-A7 cores. 
Concretely, the energy efficiency attained with the complete Cortex-A7 cluster is twice
that observed with a single core of the same type. In contrast, the best energy efficiency for the Cortex-A15
is only 33\% higher than that measured with 
a single Cortex-A15 core. Moreover, due to the non-linear increase in performance when adding
the fourth Cortex-A15 core, the most energy-efficient solution is obtained with three cores instead of the complete cluster.
It is also worth emphasizing that the exploitation of four~Cortex-A7 cores delivers significantly higher energy efficiency 
than an alternative that leverages a single Cortex-A15 core, though the overall 
performance of the former option is slightly worse.

In general, these graphs reveal that the performance achieved by the complete Cortex-A15 cluster is roughly four times
that of the Cortex-A7 cluster but their energy efficiency is similar. This last observation is interesting
since, a priori, one could expect that the Cortex-A7 cluster was more energy-efficient than the Cortex-A15 cluster.
However, we would like to remark that all our experiments report the energy efficiency of the complete SoC, 
and that the Cortex-A15 cluster in idle state already dissipates more power than a single Cortex-A7 core in execution.

\section{Architecture-Oblivious BLIS \gemm on the big.LITTLE SoC}
\label{sec:performance}

%BLIS allows to select, at run time, which 
%(one or more) of the five internal loops are parallelized. 
%Different loops can be simultaneously parallelized in
%order to adapt the execution to the specific features of the architecture.
%In particular, if one of the loops is parallelized, a static partition and 
%mapping of loop iteration chunks to the 
%OpenMP threads is performed prior to the beginning of the loop. 
%However, due to the characteristics of an AMP on which the peak performance of the cores differs
%some adaptations are required to fully exploit its capabilities.

The default approach adopted by BLIS to map \gemm on a multi-threaded CPU (see Section~\ref{Sec:Parallelization}) 
presents two main drawbacks when applied to simultaneously leverage the asymmetric cores of an AMP:

\begin{itemize}

\item BLIS relies on a static partitioning and mapping of the loop iteration space among the threads, oblivious
of the computational power of the cores these iteration chunks are assigned to.
Therefore, independently of the chunk size and the specific loops that are parallelized, this strategy can only 
yield an unbalanced distribution of the workload (basically, the micro-kernels) among the asymmetric cores.  

\item In addition, BLIS employs constant values for the loop strides that, in order to attain high performance, need to match the optimal configuration parameters 
determined by the core cache organization. However, given that we face a system with two different
architectures (Cortex-A15 and Cortex-A7), and thus different optimal cache parameters, 
we would ideally need to use different loop strides/configuration parameters for each type of core.

\end{itemize}

The following experiment 
is designed to expose the negative impact of these two mismatches between the BLIS approach and the Exynos 5422 SoC 
on the performance and energy behavior of \gemm. 
For the evaluation, 
given the guidelines in Section~\ref{Sec:Parallelization} and the lack of an L3 cache in this chip, we adopt 
the following two-level parallelization strategy:
\begin{itemize}
	%\item[Inter-cluster parallelization.] 
\item Coarse-grain (or inter-cluster): 
      Loop~1 is tackled using {\em 2-way parallelism} to statically distribute its iteration space between the 
      two clusters. %As argued earlier (see Section~\ref{Sec:Parallelization}),
      This loop (and also Loop~3) is a good candidate for parallelization across cores 
      with a proprietary and isolated L2 cache, as is the case of each cluster in the Exynos 5422 SoC.

	%\item[Intra-cluster parallelization.] 
\item Fine grain (or intra-cluster): 
      Loop~4 is parallelized using up to {\em 4-way parallelism} to statically distribute its iteration space among 
      the four cores of the same cluster. %As discussed earlier, 
      This loop (as well as Loop~5) is
      a good candidate for parallelization across cores sharing a common L2 cache, as is the case of cores in the same
      cluster of the Exynos 5422 SoC.
\end{itemize}
In addition, the cache configuration parameters are set to those that are optimal for the Cortex-A15.
We note that similar qualitative observations were obtained when parallelizing the alternative three combinations of loops
1/3 and 4/5; and/or when the cache parameters were configured using the optimal values for the Cortex-A7.  

\begin{figure}[t]
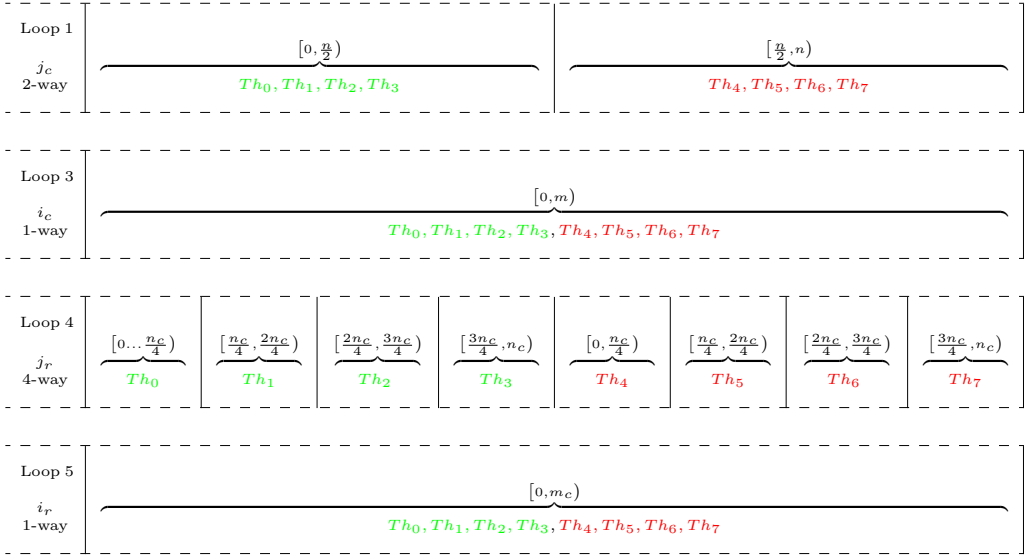


\begin{footnotesize}
\begin{tabular}{C{0.7cm} C{12.6cm}}
 & Iteration Space / Thread assignment \\
\end{tabular}
\end{footnotesize}

\bigskip

\begin{tiny}
\resizebox{\linewidth}{!}{
\begin{tabular}{C{0.7cm}| C{6.20cm} | C{6.20cm} |} 
        \hdashline
Loop~1	& & \\
	$j_c$   & \multicolumn{1}{c|}{{$\overbrace {\hspace{6.20cm}}^{\big[ 0, \frac{n}{2} \big)}$}} & \multicolumn{1}{c|}{{$\overbrace {\hspace{6.20cm}}^{\big[ \frac{n}{2}, n \big)}$}} \\ 
2-way   & $\textcolor{green}{Th_0, Th_1, Th_2, Th_3}$ & $\textcolor{red}{Th_4, Th_5, Th_6, Th_7}$\\ 
        & & \\ 
	\hdashline
\end{tabular}
}
\end{tiny}

\bigskip

\begin{tiny}
\resizebox{\linewidth}{!}{
\begin{tabular}{C{0.7cm}| C{12.8cm} |}
        \hdashline
Loop~3      & \\
	$i_c$ & \multicolumn{1}{c|}{{$\overbrace {\hspace{12.8cm}}^{\big[ 0, m \big)}$}} \\ 
1-way & \multicolumn{1}{c|}{$\textcolor{green}{Th_0, Th_1, Th_2, Th_3}, \textcolor{red}{Th_4, Th_5, Th_6, Th_7}$} \\ 
      & \\
        \hdashline
\end{tabular}
}
\end{tiny}

\bigskip

\begin{tiny}
%\begin{tabular}{C{0.7cm}| C{2.2cm} | C{2.2cm} | }
\resizebox{\linewidth}{!}{
\begin{tabular}{C{0.7cm}| C{1.2cm} | C{1.2cm} | C{1.2cm} | C{1.2cm} | C{1.2cm} | C{1.2cm} | C{1.2cm} | C{1.2cm} | }
        \hdashline
Loop~4	& & & & & & & & \\
	$j_r$   & \multicolumn{1}{c|}{{$\overbrace {\hspace{1.2cm}}^{\big[ 0              \ldots \frac{n_c}{4}\big)} $}} & \multicolumn{1}{c|}{{$\overbrace {\hspace{1.2cm}}^{\big[ \frac{n_c}{4}, \frac{2n_c}{4}\big)}$ }} 
	& \multicolumn{1}{c|}{{$\overbrace {\hspace{1.2cm}}^{\big[ \frac{2n_c}{4}, \frac{3n_c}{4} \big)}$}} & \multicolumn{1}{c|}{{$\overbrace {\hspace{1.2cm}}^{\big[ \frac{3n_c}{4}, n_c \big)             }$}}  
	& \multicolumn{1}{c|}{{$\overbrace {\hspace{1.2cm}}^{\big[ 0             , \frac{n_c}{4}  \big)} $}} & \multicolumn{1}{c|}{{$\overbrace {\hspace{1.2cm}}^{\big[ \frac{n_c}{4}, \frac{2n_c}{4} \big)}$ }} 
	& \multicolumn{1}{c|}{{$\overbrace {\hspace{1.2cm}}^{\big[ \frac{2n_c}{4}, \frac{3n_c}{4} \big)}$}} & \multicolumn{1}{c|}{{$\overbrace {\hspace{1.2cm}}^{\big[ \frac{3n_c}{4}, n_c \big)             }$}}   \\
4-way   & $\textcolor{green}{Th_0}$ & $\textcolor{green}{Th_1}$ & $\textcolor{green}{Th_2}$ & $\textcolor{green}{Th_3}$ & $\textcolor{red}{Th_4}$ & $\textcolor{red}{Th_5}$ & $\textcolor{red}{Th_6}$ & $\textcolor{red}{Th_7}$\\ 
      & & & & & & & & \\
\hdashline
%$j_r$ & $Th_0, Th_1, Th_2, Th_3$ & Th_\\ 
\end{tabular}
}
\end{tiny}

\bigskip

\begin{tiny}
\resizebox{\linewidth}{!}{
\begin{tabular}{C{0.7cm}| C{12.8cm} |}
        \hdashline
Loop~5      & \\
	$i_r$      & \multicolumn{1}{c|}{{$\overbrace {\hspace{12.8cm}}^{\big[ 0, m_c \big)}$}} \\ 
1-way																     & $\textcolor{green}{Th_0, Th_1, Th_2, Th_3}, \textcolor{red}{Th_4, Th_5, Th_6, Th_7}$ \\ 
      & \\
        \hdashline
\end{tabular}
}
\end{tiny}

\caption{Partitioning of the iteration space and assignment to threads/cores for a multi-threaded BLIS implementation 
         with 8-way parallelism that combines 2-way parallelism from Loop~1 and 4-way parallelism from~Loop 4. 
         Threads in green and red are respectively mapped to big and LITTLE cores.}
\label{fig:symmetric_iterations}
\end{figure}

Figure~\ref{fig:symmetric_iterations} illustrates 
the implications of the previous scheduling strategy in terms of loop partitioning and assignment to threads.
In total, eight threads are created and binded to the cores so that
we are extracting in overall {\em 8-way parallelism} within BLIS. 
Note how the iteration space for all loops is homogeneously distributed across the cores (i.e., without taking
into account the core type).

\begin{figure}[t]
\begin{center}
\begin{tabular}{c}
\begin{minipage}[c]{\textwidth}
\includegraphics[width=0.5\textwidth]{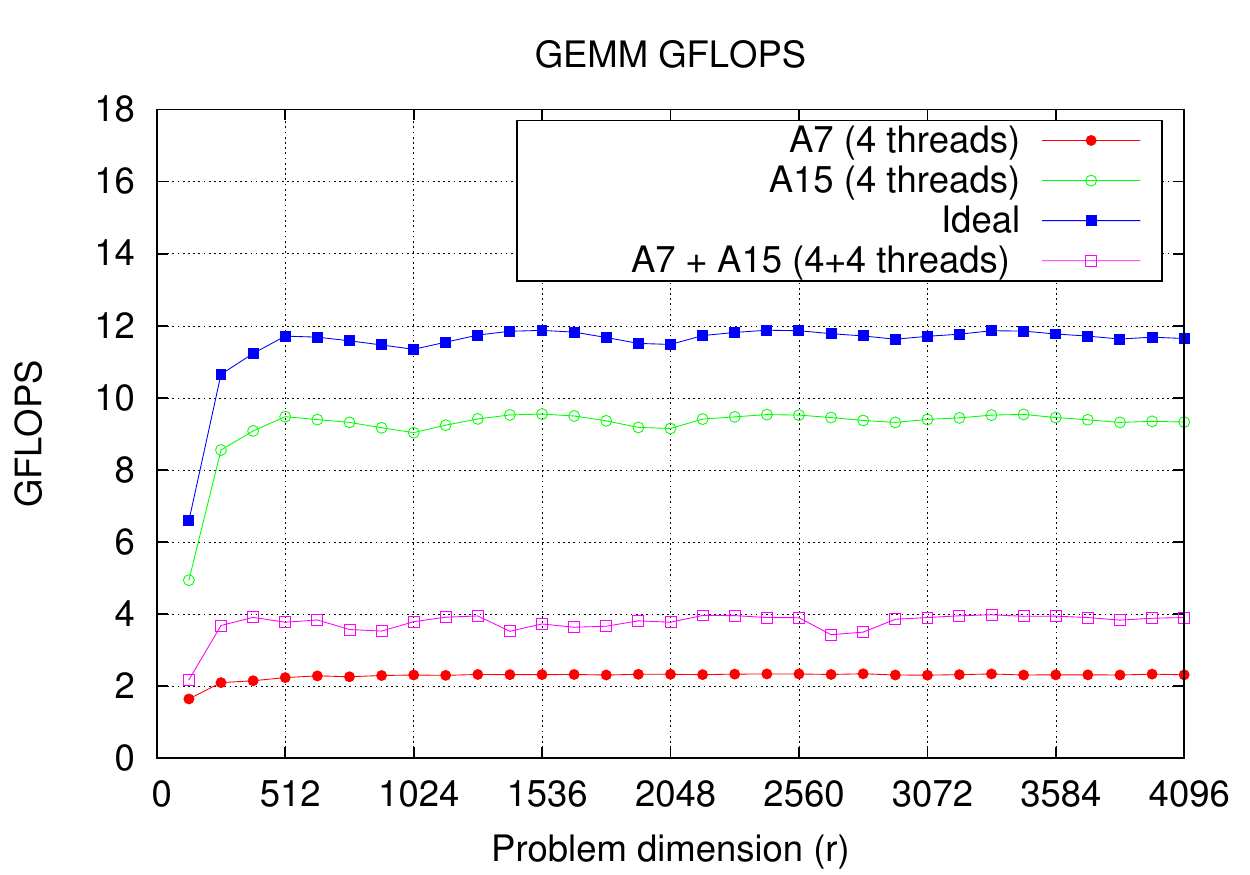}
\includegraphics[width=0.5\textwidth]{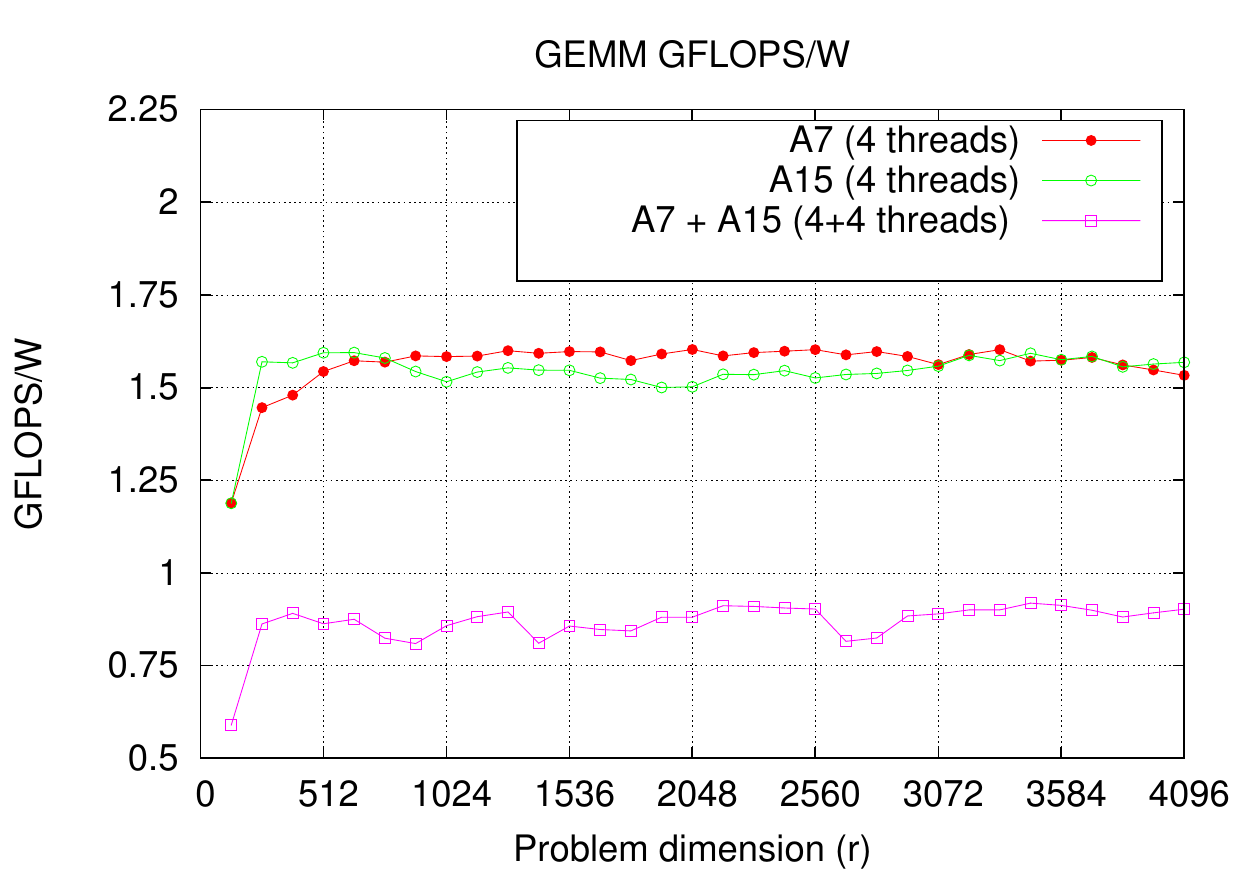}
\end{minipage}
\end{tabular}
\end{center}
\caption{Performance (left) and energy efficiency (right) of the reference BLIS \gemm using 
         exclusively one type of core in isolation, and the {\sc sss} version with a coarse-grain
         parallelization of Loop~1 and the fine-grain parallelization of Loop~4 using 4 threads
         per cluster.}
\label{fig:performance_inicial}
\end{figure}

\figurename~\ref{fig:performance_inicial} reports the performance and energy efficiency 
using the (two-level) symmetric-static scheduling ({\sc sss}) that parallelizes loops~1 and~4. 
For reference, we also include the results from the parallelization of Loop~4
that separately exploits either the four cores in the Cortex-A15 cluster or the four cores in the Cortex-A7 cluster
(see Section~\ref{sec:blis}).  
The ``{\sf Ideal}'' line in the performance graph corresponds to the aggregated performance of 
the configurations that use four cores of each of the two types in isolation (i.e., the performance of the four
Cortex-A15 cores plus the performance of the four Cortex-A7 cores). 
This is a theoretical upper bound for the performance that can be attained
when using an optimal scheduling strategy that exploits the asymmetry of the architecture.

This experiment reveals that a naive symmetric-static workload distribution, which does not consider either the differences
in the cache hierarchy between the Cortex-A15 and the Cortex-A7,  exploits the full system (8 cores) to deliver 
only about 40\% of the highest performance that is observed when employing only the four Cortex-A15 cores. 
The reason is that, with this approach, BLIS performs a static partitioning and mapping of the iteration space 
to the processing cores in a homogeneous manner.  This causes a severe workload imbalance, as the threads
running on the Cortex-A15 rapidly process their chunks, but then have to wait 
for the threads running on the slow Cortex-A7 cores to complete their work. 
The energy efficiency of the naive solution is also dramatically affected, 
and this configuration achieves the worst energy results. 
In conclusion, this experiment naturally motivates the need of
an efficient alternative to the homogeneous {\sc sss} partitioning of the iteration space integrated 
in the original multi-threaded implementation of BLIS \gemm.

\section{Architecture-Aware Optimization of BLIS \gemm on the big.LITTLE SoC}
\label{sec:strategies}

In this section, we briefly review the control mechanism that governs the parallelization of BLIS \gemm. Next, 
we propose and integrate two asymmetry-aware strategies for workload scheduling of the BLIS \gemm micro-kernels
as well as a cache-aware configuration for
AMPs; and we evaluate the impact of these techniques on performance and energy efficiency. 
The optimized implementations can be described, at a high level, as follows:
\begin{itemize}
\item {\em Static-asymmetric scheduling} ({\sc sas}).
      This option statically partitions and assigns loop iterations to different thread types
      based on the performance difference between fast and slow cores. 
\item {\em Cache-aware static-asymmetric scheduling} ({\sc ca-sas}). This strategy 
      enhances {\sc sas} by adapting the loop strides to the distinct cache configurations of the two computing clusters. 
\item {\em Cache-aware dynamic-asymmetric scheduling} ({\sc ca-das}). 
      This option improves the previous ones by replacing the static partitioning of the iteration space 
      with a dynamic workload distribution across clusters.
\end{itemize}

\subsection{BLIS internals}

The execution of all BLIS routines, including \gemm, is commanded by a {\em control-tree}. 
This is a recursive data structure that encodes all the information necessary to combine the
basic building blocks offered by the BLIS framework in order to implement high-performance algorithms for virtually any BLAS-like operation. %as depicted in Figures~\ref{fig:gotoblas_gemm} and~\ref{fig:movements_gemm}. 
The control tree for a given BLAS-3 operation governs, among others, which
combination of loops are to be executed to 
complete the operation (that is, the exact {\em algorithmic variant} to execute at each level of the
general algorithm), the loop stride for 
each loop (specific to each target architecture), and the exact points at which packing must
occur. In addition, 
for  multi-threaded BLIS implementations, 
the control tree defines which loops need to be parallelized 
and the level of concurrency to extract at each point of the algorithm.

A key property of the {\em control trees} is that they can be leveraged to modify these parameters without affecting the rest of the
BLAS implementation, boosting programmer's productivity and enhancing flexibility. In our modifications of the BLIS
framework, we have exploited this abstraction mechanism in order to encode the differences between the original framework
and our versions adapted for AMPs. 
In particular, we next focus on the necessary modifications and requirements
to implement an asymmetric scheduling of the loop iteration space to fast and slow cores, 
and the modification of the loop strides in order to 
develop a cache-aware configuration for BLIS \gemm.

\subsection{Static-asymmetric scheduling ({\sc sas})}

Taking into account the experiment in Section~\ref{sec:performance},
we have revamped the original multi-threaded implementation of BLIS \gemm to 
distinguish between the distinct computational power of the two types of cores included in the
ARM big.LITTLE architecture. 
In particular,
the {\sc sas} version of BLIS \gemm 
integrates the following two new features, which modify the behavior of the default asymmetry-oblivious 
multi-threaded implementation at execution time: 
{\em i)} a mechanism to create ``fast'' and ``slow'' threads, which are bound
upon initialization of the library to the big and LITTLE cores, respectively; and
{\em ii)} a mechanism to decide which one of the loops that are parallelized
needs to be partitioned and assigned to fast/slow cores asymmetrically.
The number of iteration chunks assigned to threads will thus no longer be the same.
Instead, these numbers will be assigned according to the capabilities of each type of core.

Our reimplementation also comprises, as an configuration knob,
an interface to specify the {\em ratio} of performance between big and LITTLE cores. 
%This configuration parameter defines the number of iterations assigned to each thread/core.
For the specific loop that is selected as candidate to partition the computational workload between the two clusters,
this configuration parameter controls the number of iteration chunks that are assigned to each cluster.
The amount of threads/cores of each type, performance ratio and specific loop to be asymmetrically
partitioned can thus be modified via ad-hoc environment variables, and they can all be fixed at execution time 
in order to tune the behavior
of the library to other specific big.LITTLE setups (for example, to changes in the core frequency that affect
the performance ratio between core types).

This new functionality is fully configurable and has been embedded into the internal control tree structures that
govern the parallelization of each loop in the general BLIS \gemm algorithm.

%{\bf This option is architecture-oblivous but... ARE WE USING DIFFERENT MICRO-KERNELS FOR THE A7 and A15? We had to modify the control-tree for that, correct? Should this then be in the next subsection? Rafa: No, de momento solo se reparte de manera asimetrica}

\subsubsection{Mapping the iteration space to clusters and cores}
\label{subsubsec:coarsevsfine}

\begin{figure}[t]
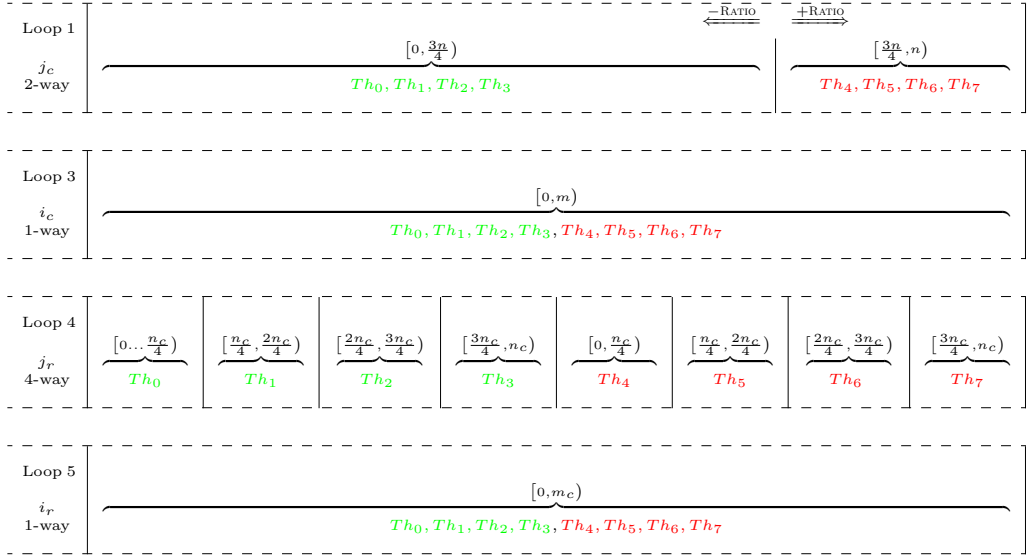


\begin{footnotesize}
\begin{tabular}{C{0.7cm} C{12.6cm}}
 & Iteration Space / Thread assignment \\
\end{tabular}
\end{footnotesize}

\bigskip

\begin{tiny}
\resizebox{\linewidth}{!}{
\begin{tabular}{C{0.7cm}| C{9.30cm} | C{3.10cm} |} 
        \hdashline
	%%& & \\
Loop~1	&                                                                       \multicolumn{1}{ r }{$ \xLeftarrow{- \textsc{Ratio}} $} & \multicolumn{1}{ l|}{$ \xRightarrow{+ \textsc{Ratio}} $}                                                                \\ 
	$j_c$   & \multicolumn{1}{c|}{{$\overbrace {\hspace{ 9.30cm}}^{\big[ 0, \frac{3n}{4} \big)}$}} & \multicolumn{1}{c|}{{$\overbrace {\hspace{3.10cm}}^{\big[\frac{3n}{4}, n \big)}$}} \\ 
2-way   & $\textcolor{green}{Th_0, Th_1, Th_2, Th_3}$ & $\textcolor{red}{Th_4, Th_5, Th_6, Th_7}$\\ 
        & & \\ 
	\hdashline
\end{tabular}
}
\end{tiny}

\bigskip

\begin{tiny}
\resizebox{\linewidth}{!}{
\begin{tabular}{C{0.7cm}| C{12.8cm} |}
        \hdashline
Loop~3      & \\
	$i_c$ & \multicolumn{1}{c|}{{$\overbrace {\hspace{12.8cm}}^{\big[ 0, m \big)}$}} \\ 
1-way & \multicolumn{1}{c|}{$\textcolor{green}{Th_0, Th_1, Th_2, Th_3}, \textcolor{red}{Th_4, Th_5, Th_6, Th_7}$} \\ 
      & \\
        \hdashline
\end{tabular}
}
\end{tiny}

\bigskip

\begin{tiny}
%\begin{tabular}{C{0.7cm}| C{2.2cm} | C{2.2cm} | }
\resizebox{\linewidth}{!}{
\begin{tabular}{C{0.7cm}| C{1.2cm} | C{1.2cm} | C{1.2cm} | C{1.2cm} | C{1.2cm} | C{1.2cm} | C{1.2cm} | C{1.2cm} | }
        \hdashline
Loop~4	& & & & & & & & \\
	$j_r$   & \multicolumn{1}{c|}{{$\overbrace {\hspace{1.2cm}}^{\big[ 0              \ldots \frac{n_c}{4}\big)} $}} & \multicolumn{1}{c|}{{$\overbrace {\hspace{1.2cm}}^{\big[ \frac{n_c}{4}, \frac{2n_c}{4}\big)}$ }} 
	& \multicolumn{1}{c|}{{$\overbrace {\hspace{1.2cm}}^{\big[ \frac{2n_c}{4}, \frac{3n_c}{4} \big)}$}} & \multicolumn{1}{c|}{{$\overbrace {\hspace{1.2cm}}^{\big[ \frac{3n_c}{4}, n_c \big)             }$}}  
	& \multicolumn{1}{c|}{{$\overbrace {\hspace{1.2cm}}^{\big[ 0             , \frac{n_c}{4}  \big)} $}} & \multicolumn{1}{c|}{{$\overbrace {\hspace{1.2cm}}^{\big[ \frac{n_c}{4}, \frac{2n_c}{4} \big)}$ }} 
	& \multicolumn{1}{c|}{{$\overbrace {\hspace{1.2cm}}^{\big[ \frac{2n_c}{4}, \frac{3n_c}{4} \big)}$}} & \multicolumn{1}{c|}{{$\overbrace {\hspace{1.2cm}}^{\big[ \frac{3n_c}{4}, n_c \big)             }$}}   \\
4-way   & $\textcolor{green}{Th_0}$ & $\textcolor{green}{Th_1}$ & $\textcolor{green}{Th_2}$ & $\textcolor{green}{Th_3}$ & $\textcolor{red}{Th_4}$ & $\textcolor{red}{Th_5}$ & $\textcolor{red}{Th_6}$ & $\textcolor{red}{Th_7}$\\ 
      & & & & & & & & \\
\hdashline
%$j_r$ & $Th_0, Th_1, Th_2, Th_3$ & Th_\\ 
\end{tabular}
}
\end{tiny}

\bigskip

\begin{tiny}
\resizebox{\linewidth}{!}{
\begin{tabular}{C{0.7cm}| C{12.8cm} |}
        \hdashline
Loop~5      & \\
	$i_r$      & \multicolumn{1}{c|}{{$\overbrace {\hspace{12.8cm}}^{\big[ 0, m_c \big)}$}} \\ 
1-way																     & $\textcolor{green}{Th_0, Th_1, Th_2, Th_3}, \textcolor{red}{Th_4, Th_5, Th_6, Th_7}$ \\ 
      & \\
        \hdashline
\end{tabular}
}
\end{tiny}

\caption{Partitioning of the iteration space and assignment to threads/cores for a multi-threaded BLIS implementation 
with 8-way parallelism that asymmetrically combines 2-way parallelism from Loop 1 (using a ratio between fast and slow cores of 3) 
and 4-way parallelism from Loop 4.}
\label{fig:asymmetric_iterations}
\end{figure}

Given the memory organization of the Exynos 5422 SoC, and the guidelines given for the parallelization of BLIS
\gemm in section~\ref{sec:blis}, we evaluated the following parallelization options for {\sc sas}:

\begin{itemize}
  \item Coarse-grain: the micro-kernels of the complete multiplication %$C \mathrel{+}= A \cdot B$
are distributed among the Cortex-A15 and Cortex-A7 clusters by parallelizing
either Loop~1 or Loop~3, 
%%% A distribution of the workload in the two inner-most loops (i.e., 4 or 5)
%%% is not considered since this would %require separate memory buffers to hold $A_c$ for both types of threads,
%%% duplicate  $A_c$ in the L2 cache of both clusters.      
%%% In order to %preserve the optimal cache parameters during the execution of \gemm, while 
%%% attain a distribution of the workload proportional to the computational
%%% power of the Cortex-A15 vs Cortex-A7 clusters, we assign
with a different number of iterations of the parallelized loop assigned to each cluster ({\em 2-way parallelism}).
%%% see Figure~\ref{fig:A15vsA7}.
%In particular, the ratio applied to distribute the iteration space
%between the Cortex-A15 and Cortex-A7 for \gemm\ has been empirically determined to be 6:1\footnote{This ratio varies depending on the target architecture, 
%core operating frequency, and specific routine, so it should be adjusted accordingly.}.
%%% The distribution ratio applied to partition the iteration space of \gemm
%%% between the Cortex-A15 and Cortex-A7 clusters is empirically determined, as it depends on the target architecture
%%% and core operating frequency.

\item Fine-grain: the execution of each macro-kernel 
%%% ($\textcolor{darkgreen}{C_c} \mathrel{+}= \textcolor{darkred}{A_c} \cdot \textcolor{darkblue}{B_c}$, see Figure~\ref{fig:gotoblas_gemm})
is partitioned among the cores of the same 
type by parallelizing Loop~4, Loop~5, or both ({\em 4-way parallelism}). %%% ; see Figure~\ref{fig:Cores4}.

\end{itemize}
To illustrate this, Figure~\ref{fig:asymmetric_iterations}
depicts the distribution of the iteration space across fast and slow threads for an scenario in which the
iteration space of Loop~1 is asymmetrically distributed across fast and slow threads, using a ratio 3, 
so that the fast threads are assigned three times 
%as many iterations of that loop 
more computations
than the slow threads.
Internally, Loop~4 is parallelized to distribute the work among the cores of the same cluster.

%\end{itemize}

%%% \begin{figure}[t]
%%% \begin{center}
%%% \includegraphics[height=5cm]{Figures/A15vsA7.pdf}
%%% \end{center}
%%% \caption{\label{fig:A15vsA7} Workload distributions 
         %%% for the matrix multiplication $C \mathrel{+}= A \cdot B$
         %%% between the A15 and A7 quad-core clusters. 
         %%% Top: parallelization of Loop~1 ($j_c$); 
         %%% bottom: parallelization of Loop~3 ($i_c$).
         %%% In the bottom plot, the small rectangles, delimited by the fine lines,
         %%% denote the operands of the macro-kernel 
         %%% $\textcolor{darkgreen}{C_c} \mathrel{+}= \textcolor{darkred}{A_c} \cdot 
         %%% \textcolor{darkblue}{B_c}$.}
         %$C \mathrel{+}= A \times B$, with
         %%$m \times n$ $C$, $m \times k$ $A$ and $k \times n$ $B$.  
%%% \end{figure}

%%% \begin{figure}[t]
%%% \begin{center}
%%% \includegraphics[height=5cm]{Figures/Cores4.pdf}
%%% \end{center}
%%% \caption{\label{fig:Cores4} Workload distributions 
         %%% for the macro-kernel multiplication 
         %%% $\textcolor{darkgreen}{C_c} \mathrel{+}= \textcolor{darkred}{A_c} \cdot 
         %%% \textcolor{darkblue}{B_c}$
         %%% between four cores of the same type 
         %%% (C$_{\mbox{\rm 0}}$, C$_{\mbox{\rm 1}}$, C$_{\mbox{\rm 2}}$, C$_{\mbox{\rm 3}}$). 
         %%% Top: parallelization of Loop~4 ($j_r$); 
         %%% bottom: parallelization of Loop~5 ($i_r$). In this example,
         %%% the OpenMP chunk size equals~2 in the first case and~4 in the second.}
%%% \end{figure}

\subsubsection{Evaluation of {\sc sas}}

The combination of the coarse-grain and fine-grain 
parallelization strategies for {\sc sas} %%% illustrated in Figures~\ref{fig:A15vsA7} and~\ref{fig:Cores4} 
yields four direct parallelization schemes. Additionally, two more configurations are possible, 
combining the coarse-grain parallelization
of either Loop~1 or Loop~3 with the fine-grain parallelization of both Loops~4 and~5.
%However, for brevity, we only report results for the former four, using (distribution) ratios for the inter-cluster parallelization that range from~3 to~7.
For brevity, because the qualitative conclusions that can be extracted from these parallelization strategies are very
similar, 
we only report results when the iteration space is distributed between the clusters
in Loop~1; and the macro-kernel is partitioned among homogeneous cores in Loop~4, using 
(distribution) ratios for the coarse-grain parallelization that range from~1 to~7.

\begin{figure}[t]
\begin{center}
\begin{tabular}{c}
\begin{minipage}[c]{\textwidth}
\includegraphics[width=0.5\textwidth]{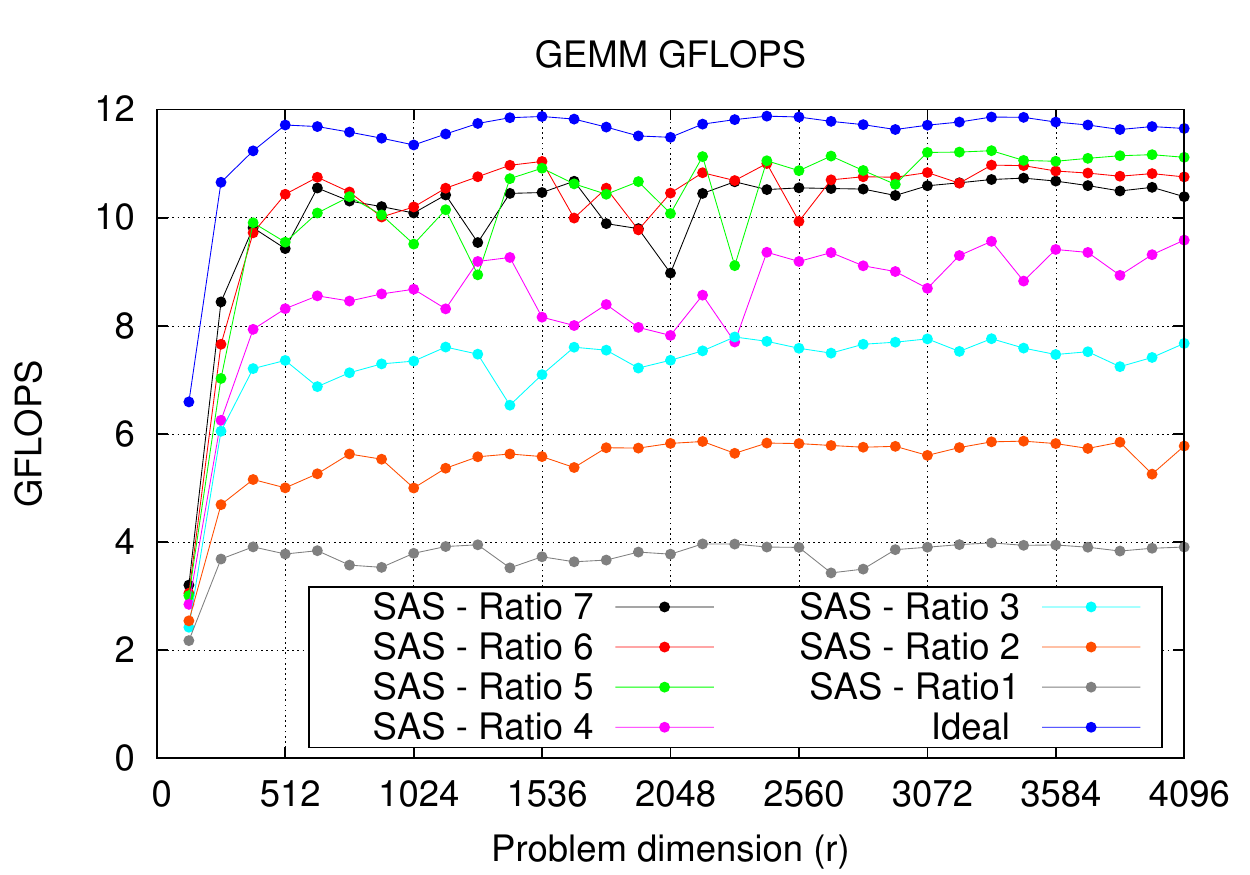}
\includegraphics[width=0.5\textwidth]{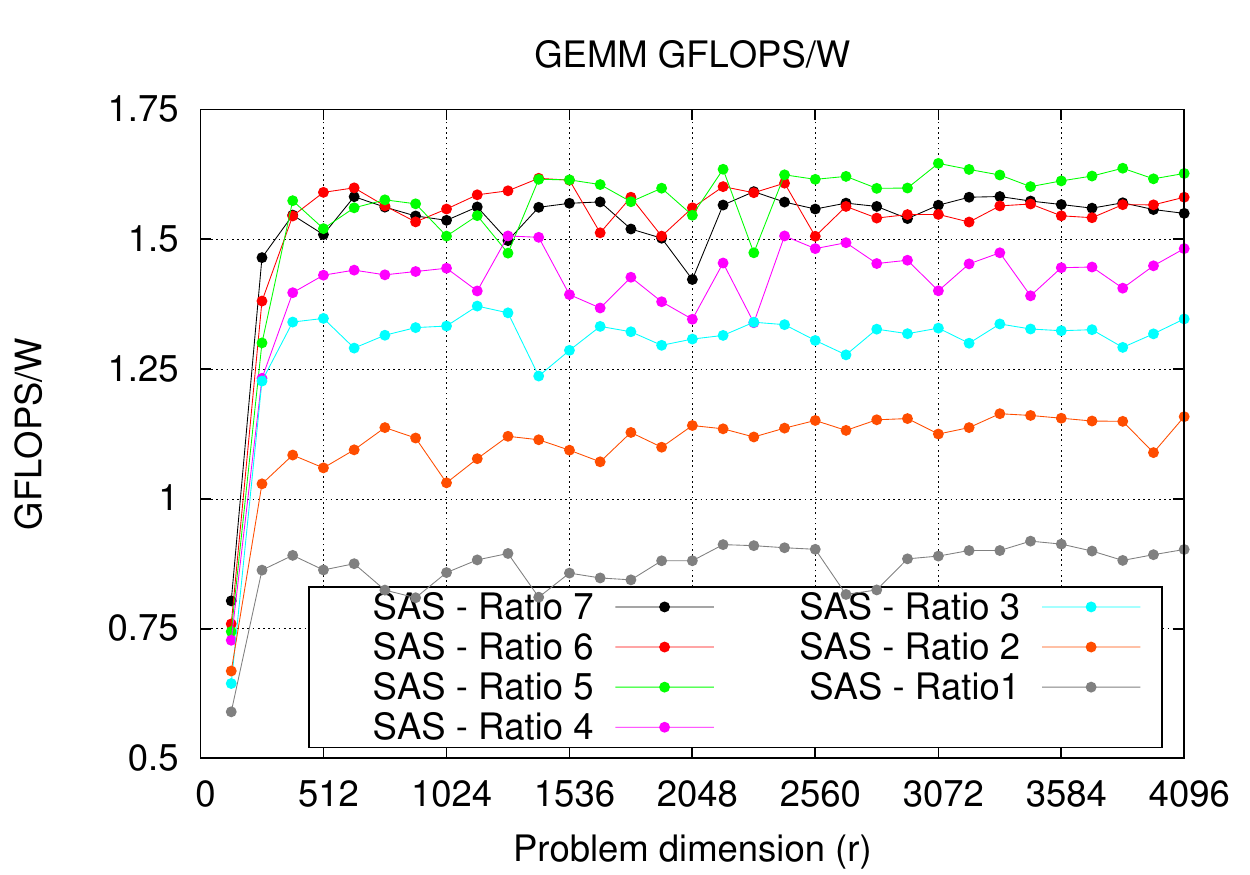}
\end{minipage}
\end{tabular}
\end{center}
\caption{Performance (left) and energy-efficiency (right) of the {\sc sas} version
of BLIS \gemm with a coarse-grain parallelization of Loop~1 and a fine-grain parallelization of Loop~4 using 4~threads per
cluster.}
\label{fig:1-4_static}
\end{figure} 

%Figure~\ref{fig:AMP} reports the results for this second evaluation.
%The line labeled as ``{big.LITTLE (4+4 threads)}'' 
%corresponds to the asymmetric-aware implementation. 
%The same \gemm kernels were computed with BLIS using a symmetric workload distribution 
%(the iteration space is equally distributed among the Cortex-A7 and Cortex-A15 cores), 
%with the results labelled as ``{A7+A15 (4+4 threads)}'' in the figure. 
%For comparison purposes, the performance and energy obtained using exclusively 
%four Cortex-A7 or four Cortex-A15 CPUs are added, as well as, the ``ideal'' case lines
%corresponding to the sum of the peak performances of the configurations 
%that use four cores of each of the two types in isolation
% (i.e., the performance of the four
%Cortex-A15 cores plus the performance of the four Cortex-A7 cores).

\figurename~\ref{fig:1-4_static} displays the results for this experiment. %with 
%%% the ``{\sf Ideal}'' case added in the performance plots for reference.
The performance results show that, %, regardless of the parallelized loops, 
when the appropriate workload distribution is applied,
the asymmetric-aware {\sc sas} outperforms the peak performance of all other configurations,
being close to that of the ideal case. 
In particular, the left-hand side graph reveals that the worst performance is achieved when the ratio is~1 
(i.e., an homogeneous inter-cluster prallelization). Also, %3 
%(even worse performance was obtaine with the ratio~1, i.e., a symmetric workload distribution), 
the performance grows until a ratio of 5--6 is used, 
and above 6, in general declines with a lower limit existing at 
the performance line delivered by the Cortex-A15 cluster in isolation (not included in the figure for clarity).
These results indicate that ratios below~5 map that too much workload 
to the Cortex-A7 cluster, and ratios above~6 assign an excessive workload to the Cortex-A15 cluster.
%Finally, is worth emphasizing that the fine-grain parallelization of Loop~4 yields performance figures
%closer to those of the ideal case than the alternative that parallelizes Loop~5. The reason is that $n_c$ (linked to Loop~4)
%is usually much larger than $m_c$ (linked to Loop~5) and, therefore, it is easier to attain a more balanced workload distribution.
%(see Section~\ref{sec:blis}).
 
For the largest tested problem, the increment of performance for {\sc sas}
 compared with the configuration that employs four Cortex-A15 cores only is close to 20\%. 
However, {\sc sas} offers lower performance for the small problems,
as the chunks assigned to the big and LITTLE cores are, in those cases, too small to exploit the asymmetric architecture.
%Note that, the workload distribution is set a priori taking into account the CPU's peak performance and not the \gemm performance. 

In terms of energy efficiency, 
when the appropriate workload distribution is in place,
{\sc sas} delivers the same flops per Joule as the setup that exclusively employs the Cortex-A15 cluster.
On the other hand, when the workload is unbalanced, the energy performance is greatly affected,
as the fast threads remain idle but active, polling and consuming energy, till the slow threads complete their work.

\subsection{Cache-aware static-asymmetric scheduling ({\sc ca-sas})}

The original implementation of BLIS contains a single control-tree per operation, which implies that the \gemm routine
can only employ using the optimal cache configuration parameters for either the Cortex-A15 or the Cortex-A7.
Our solution to this problem duplicates the control structure to set different configuration values
for $m_c$ and $k_c$, depending on the type of core. Specifically, two different control-trees
are created upon initialization, for ``fast'' and ``slow'' threads, each setting the optimal 
loop strides/cache parameters for a different core architecture (see Section~\ref{sec:blis}).
In addition, this mechanism opens the door to the use of
specific highly-tuned micro-kernels adapted to each micro-architecture in the AMP
(and, therefore, optimal values for $m_r$ and $n_r$), depending on the type of core that executes it. 
We note that, as argued earlier in Section \ref{sec:blis}, the performance of \gemm is quite independent of $n_c$,
since there is not a L3 cache in the Exynos 5422 SoC. Furthermore, we leverage the same micro-kernel for both the 
Cortex-A7 and Cortex-A15 clusters since, in this particular SoC, it is optimal for both. 

An important caveat of this approach is that there may be some dependencies between the optimal configurations 
used for the clusters. Concretely, if the micro-kernels  
are distributed among the Cortex-A15 and Cortex-A7 clusters by parallelizing
Loop~1, independent buffers are used for $A_c$ and $B_c$, and no dependencies arise. 
However, if they are partitioned between the clusters by parallelizing Loop~3, then the buffer for $B_c$ is shared, and  
it is necessary to employ a common value of $k_c$ for the Cortex-A15 and the Cortex-A7. 
%In consequence, when Loop~3 is used to distribute the computational workload between the cluster, 
In this scenario the parameter is set to $k_c = 952$ in both control-trees, and a new (sub)optimal value
for $m_c$ has to be obtained for the Cortex-A7 threads. 
In order to do that, we carried out a similar search as that exposed in Section~\ref{sec:blis}, finding the new optimal value at 
$m_c = 32$ for the Cortex-A7 (taking into account that the $k_c$ parameter depends on the Cortex-A15). 
With these new optimal parameters, the performance peak attained with the Cortex-A7 cluster
is slightly worse than that observed the actual Cortex-A7-specific optimal parameters. 
However, it is still higher than that obtained with the
cache parameters for the Cortex-A15 as, with those much larger values, the memory buffer $A_c$ does not fit into the 
small L2 cache of the Cortex-A7. 

%In this round of experiments we evaluate the performance and energy-efficiency of the 
%architecture-oblivious port of BLIS to the big.LITTLE architecture. For this purpose,
%we run a collection of \gemm kernels, 
%relaying on a 2-way parallelization to distribute iterations of Loop~1 
%(\figurename~\ref{fig:1-4-ASYM} and \figurename~\ref{fig:1-5-ASYM})
%or to distribute iterations of Loop~3 (\figurename~\ref{fig:3-4-ASYM} and \figurename~\ref{fig:3-5-ASYM})
%%(see Section \ref{sec:performance}),
%with a ratios of 1, 3, and 5, among the cores of the fast and slow clusters, taking advantage of
%the independent L2 cache per cluster in this manner.  
%For the fine-grain parallelization, four threads are leveraged 
%in order to assign chunks of the iteration space 
%for Loop~4 (\figurename~\ref{fig:1-4-ASYM} and \figurename~\ref{fig:3-4-ASYM}) 
%or for Loop~5 (\figurename~\ref{fig:1-5-ASYM} and \figurename~\ref{fig:3-5-ASYM}) 
%to each core within the cluster. 
%For comparison purposes, the performance and energy-efficiency obtained using exclusively 
%four Cortex-A7 or four Cortex-A15 core are also added, as well as, the ``ideal'' case lines
%corresponding to the sum of the peak performances of the configurations 
%that use four cores of each of the two types in isolation.

%{\bf Comment on results! What about the use of architecture-specific micro-kernels? Is it done at this point, as part of the modification
%of the control tree, or was it done already in the previous section? RAFA:CREO QUE EN NINGUNA}

\subsubsection{Comparison of {\sc sas} and {\sc ca-sas}}

The combination of the coarse-grain and fine-grain parallelization strategies described in 
Section~\ref{subsubsec:coarsevsfine} 
yields the same parallelization options for {\sc ca-sas}. For the same reasons, we only report next the results 
corresponding to an scenario where the iteration space is distributed between the clusters across Loop~1, while the macro-kernel is partitioned within
clusters in Loop~4, using (distribution) ratios for the inter-cluster parallelization of~1, 3 and~5.
For each distribution ratio, we include two lines, corresponding to the use of two control-trees ({\sc ca-sas}) and 
only one ({\sc sas}). 
%For comparison purposes, the ``ideal'' case line
%corresponding to the sum of the peak performances of the configurations 
%that use four cores of each of the two types in isolation is also added.

\begin{figure}[t]
\begin{center}
\begin{tabular}{c}
\begin{minipage}[c]{\textwidth}
\includegraphics[width=0.5\textwidth]{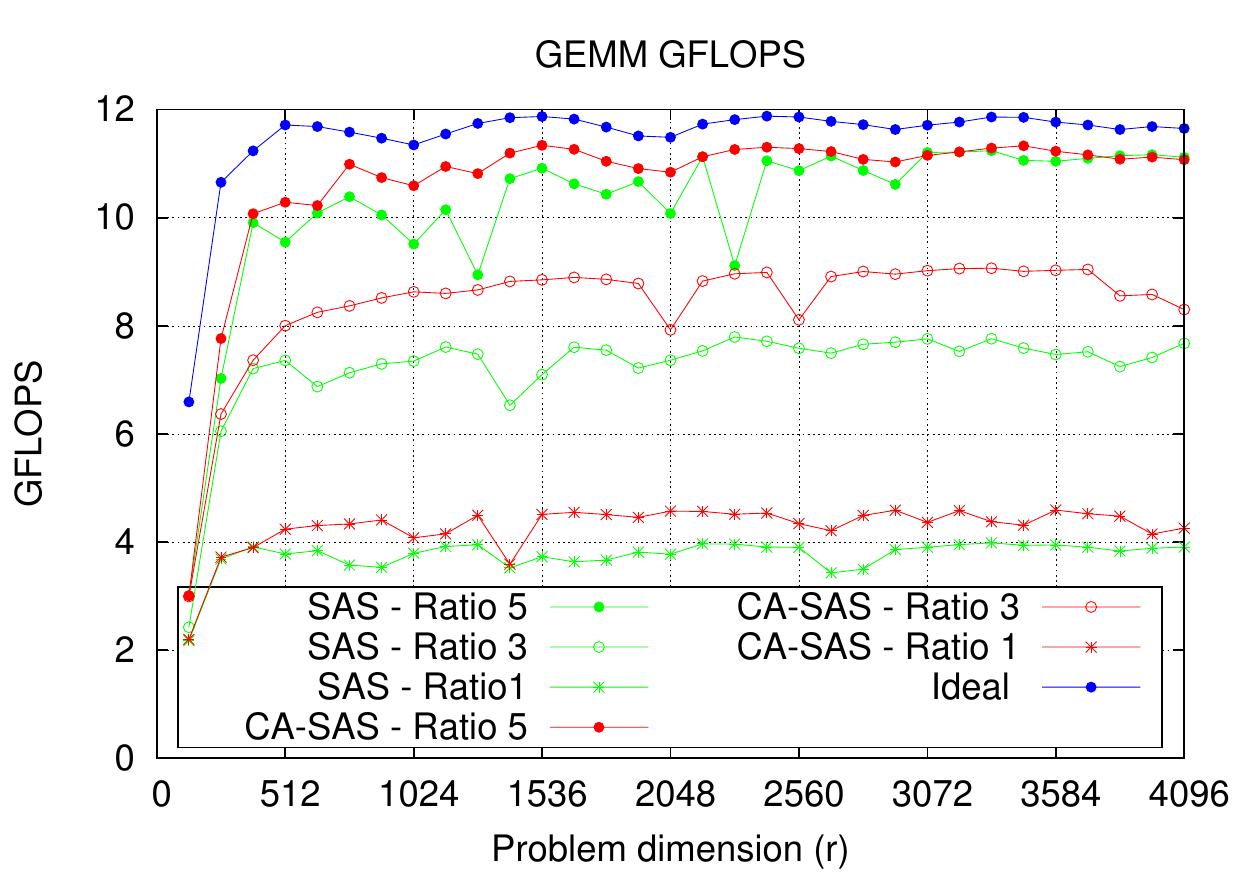}
\includegraphics[width=0.5\textwidth]{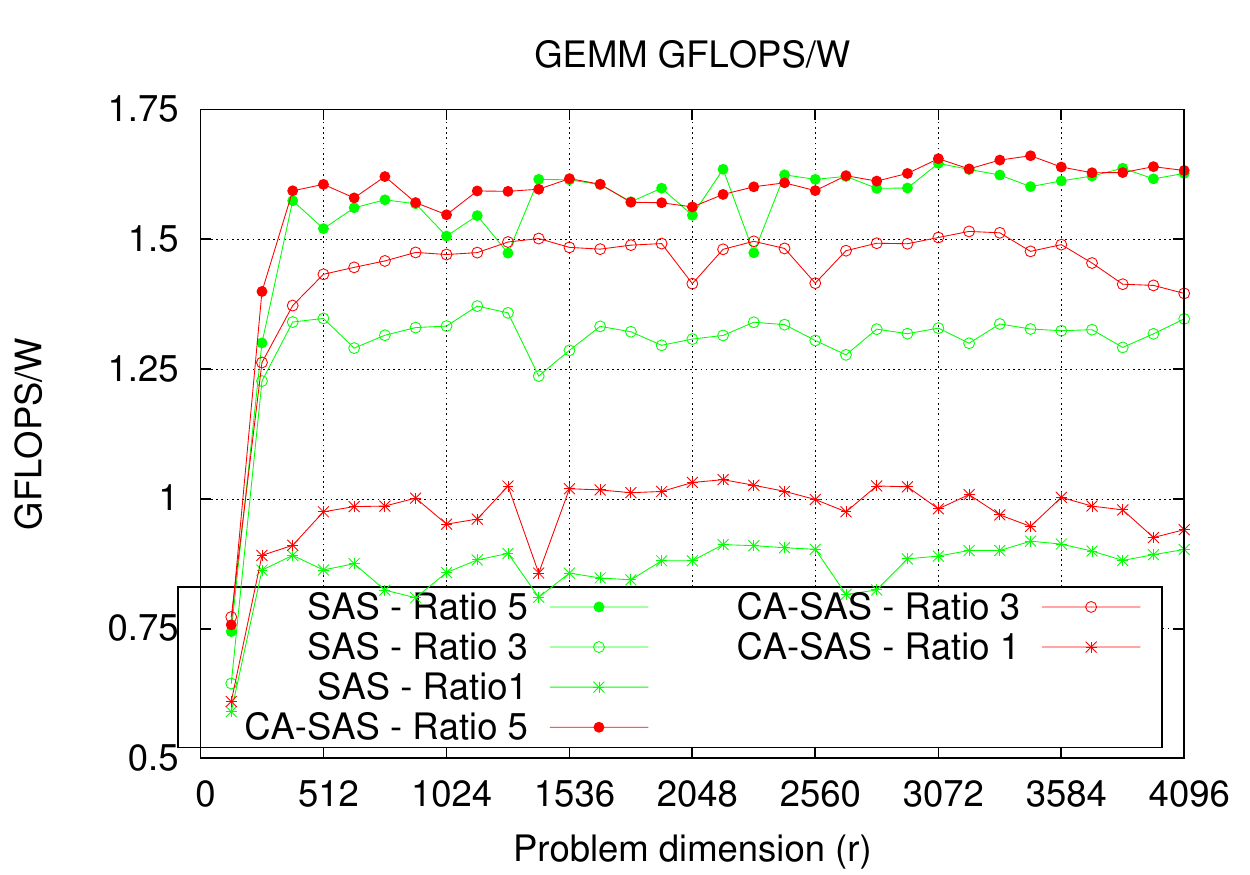}\\
%\end{minipage}
%\end{tabular}
%\end{center}
%\caption{Performance (left) and energy-efficiency (right) of the BLIS DGEMM 
         %distributing iterations of Loop~1 and four thread leveraged in Loop~4.}
%\label{fig:1-4-ASYM}
%\end{figure} 

%\begin{figure}[!t]
%\begin{center}
%\begin{tabular}{c}
%\begin{minipage}[c]{\textwidth}
%\includegraphics[height=3.7cm,width=0.5\textwidth]{./Graph/GFLOPS1-5.pdf}
%\includegraphics[height=3.7cm,width=0.5\textwidth]{./Graph/GFLOPSW1-5.pdf}\\
%\end{minipage}
%\end{tabular}
%\end{center}
%\caption{Performance (left) and energy-efficiency (right) of the BLIS DGEMM 
         %distributing iterations of Loop~1 and four thread leveraged in Loop~5.}
%\label{fig:1-5-ASYM}
%\end{figure} 

%\begin{figure}[!t]
%\begin{center}
%\begin{tabular}{c}
%\begin{minipage}[c]{\textwidth}
%\includegraphics[height=3.7cm,width=0.5\textwidth]{./Graph/GFLOPS3-4.pdf}
%\includegraphics[height=3.7cm,width=0.5\textwidth]{./Graph/GFLOPSW3-4.pdf}\\
%\end{minipage}
%\end{tabular}
%\end{center}
%\caption{Performance (left) and energy-efficiency (right) of the BLIS DGEMM 
         %distributing iterations of Loop~3 and four thread leveraged in Loop~4.}
%\label{fig:3-4-ASYM}
%\end{figure} 

%\begin{figure}[!t]
%\begin{center}
%\begin{tabular}{c}
%\begin{minipage}[c]{\textwidth}
%\includegraphics[height=3.7cm,width=0.5\textwidth]{./Graph/GFLOPS3-5.pdf}
%\includegraphics[height=3.7cm,width=0.5\textwidth]{./Graph/GFLOPSW3-5.pdf}
\end{minipage}
\end{tabular}
\end{center}
\caption{Performance (left) and energy-efficiency (right) of the 
	{\sc sas} and {\sc ca-sas} 
         versions of BLIS \gemm 
         with a coarse-grain parallelization of Loop~1 and a fine-grain parallelization of Loop~4 using
         4 threads per cluster.}
%\caption{%Performance (left) and energy-efficiency (right) of the BLIS DGEMM 
%         %distributing iterations of Loop~3 and four thread leveraged in Loop~5.}
%         Performance (left) and energy-efficiency (right) of the BLIS \gemm with static-asymmetric
%         scheduling and cache-aware configuration, from top to bottom parallelizing loops 1+4, 1+5, 3+4, and 3+5.}
%\label{fig:3-5-ASYM}
\label{fig:1-4_ASYM}
\end{figure}

The plots in \figurename~\ref{fig:1-4_ASYM} illustrate that, for both metrics of interest, better results are 
obtained with the option that integrates two control-trees. 
The increases of performance and energy efficiency are a direct consequence of 
the accelerated execution of the workload assigned to the Cortex-A7 cluster, derived from the use of more convenient
cache configuration parameters. 
We notice that the improvements at this point are only visible when too much 
work is assigned to the Cortex-A7 cluster (i.e., for distribution ratios below 5). However, as we will expose later,
this strategy has a more visible impact when a dynamic workload distribution is adopted.

\begin{figure}[]
\begin{center}
\begin{tabular}{c}
\begin{minipage}[c]{\textwidth}
\includegraphics[width=0.5\textwidth]{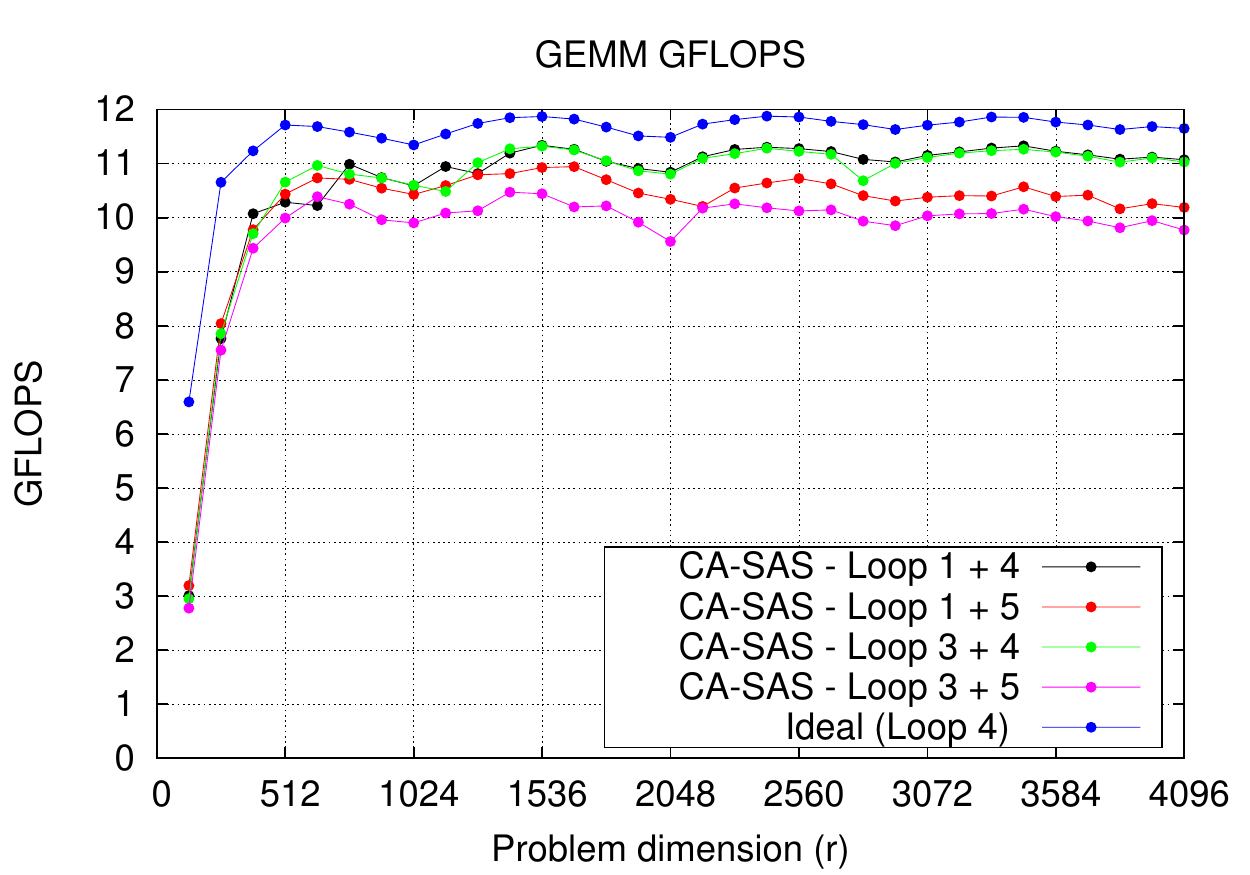}
\includegraphics[width=0.5\textwidth]{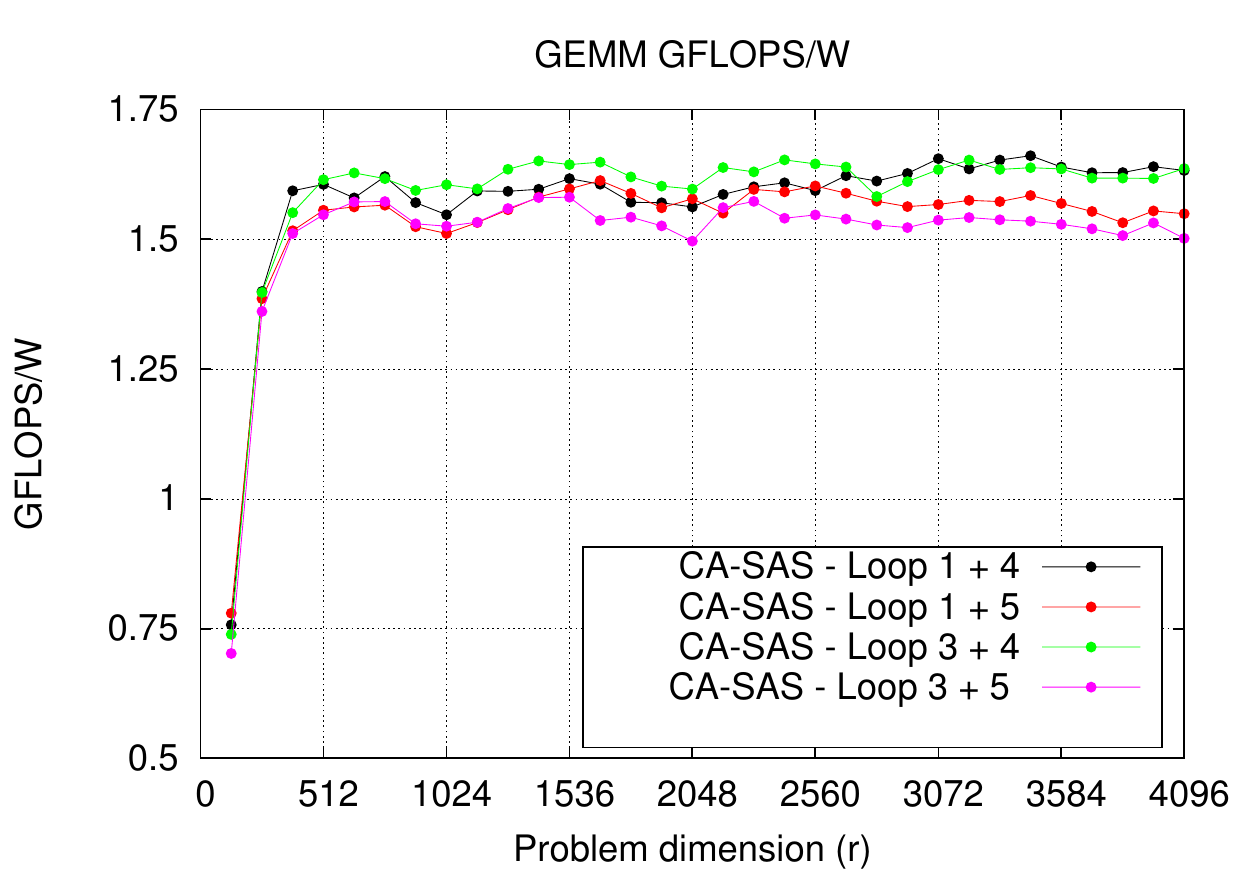}
\end{minipage}
\end{tabular}
\end{center}
\caption{Performance (left) and energy-efficiency (right) of the {\sc ca-sas} version of BLIS \gemm 
         with a coarse-grain parallelization of either Loop~1 or Loop~3; combined with a fine-grain
         parallelization of either Loop~4 or Loop 5, using a ratio 5 in both cases and 4 threads per cluster.}
%\caption{%Performance (left) and energy-efficiency (right) of the BLIS DGEMM 
%         %distributing iterations of Loop~3 and four thread leveraged in Loop~5.}
%         Performance (left) and energy-efficiency (right) of the BLIS \gemm with static-asymmetric
%         scheduling and cache-aware configuration, from top to bottom parallelizing loops 1+4, 1+5, 3+4, and 3+5.}
%\label{fig:3-5-ASYM}
\label{fig:Ratio5_ASYM}
\end{figure}

To conclude the evaluation of the {\sc ca-sas} implementation of BLIS, we compare 
the four direct combinations (parallelization options) of the coarse-grain (Loop~1 or Loop~3) and fine-grain (Loop~4 or Loop~5) options, 
%parallelization strategies illustrated in Figures~\ref{fig:A15vsA7} and~\ref{fig:Cores4} for a 
for a concrete distribution ratio of 5, using two control-trees.
\figurename~\ref{fig:Ratio5_ASYM} reports the outcome from this evaluation. 
%with the ``{\sf Ideal}'' case added again in the performance plots for reference. 
The plots show that the fine-grain parallelization of Loop~4 yields performance curves
closer to that of the ideal case than the alternatives that parallelize Loop~5. The reason is 
that $n_c$ (linked to Loop~4)
is usually much larger than $m_c$ (linked to Loop~5) and, therefore, it is easier to attain a 
more balanced workload distribution with this option. Although it is not possible to leverage 
the actual optimal cache parameters specific to the Cortex-A7 cluster when Loop~3 is parallelized the plots also reveal that,
when the fine-grain parallelization is set Loop~4, there is no noticeable difference between distributing the 
computational workload in either Loop~1 or in Loop~3; however the difference is present when the fine-grain parallelization is set in Loop~5.

\subsection{Cache-aware dynamic-asymmetric scheduling ({\sc ca-das})}

Our final step towards attaining a high performance implementation of BLIS \gemm for an AMP SoC 
integrates a mechanism that dynamically distributes the workload between the two SoC clusters. 
The main advantage of this option is that a
predefined distribution ratio becomes unnecessary, though the target loop this feature is applied to
still needs to be chosen with care.

The candidates to apply a dynamic distribution are, obviously, Loop~1 and Loop~3, since these have been previously identified 
as the best options to distribute the computational workload between the two clusters. However, the 
cache parameter $n_c$ (linked to the stride of Loop~1) is often in the order of several hundreds up 
to a few thousands and, therefore, in practice it is too large to dynamically distribute the iteration space. 
In contrast, the cache parameter $m_c$ (linked to the stride of Loop~3) is usually 
in the order of a few hundreds, and thus it is a good candidate to dynamically distribute the iterations. 
Diving into details, $n_c=4,096$ for both types of cores, while 
$m_c=32$ and $152$ for the Cortex-A7 and Cortex-A15 cores, respectively. 
In consequence, the coarse-grain dynamic distribution of the workload will be carried out across Loop~3, with 
two independent control-trees in place binded to ``fast'' and ``slow'' threads.
Note that, like in the %static {\sc sas} and 
{\sc ca-sas} scheduling strategy, the buffer $B_c$ is shared by both clusters 
and, in consequence, $k_c$ is set to $952$ for both types of cores (cache-aware optimization). 
 
The application of a dynamic scheduling strategy removes the static partitioning carried out before Loop~3.
This is replaced by a mechanism where, at each iteration of Loop~3, a single
thread bound to a ``fast'' core and a single thread bound to a ``slow'' core select the current chunk size,
which depend on the configuration parameter $m_c$ of each type of core. 
The selected workload is broadcast to the remaining threads of the same type. 
The fine-grain parallelization remains unmodified and targets Loop~4, Loop~5 or both.
The chunk size selection is performed inside a critical section that controls the execution of Loop~3. 
%a new thread synchronization point is needed to implement this feature. 
The overhead of this synchronization point %and the critical section overhead
is fully amortized 
by the utilization of a more flexible workload distribution. %regardless the target architecture
%and core operating frequency.

\subsubsection{Evaluation of {\sc ca-das}}

This last round of experiments presents a more reduced number of options, since Loop~1 
was identified as a poor choice to dynamically distributing the computational workload.
We report results when the iteration space is dynamically distributed between clusters across Loop~3, and the macro-kernel 
is partitioned within clusters in Loop~4 or in  Loop~5, using either two control-trees (one for ``fast'' 
and one for ``slow'' threads, {\sc ca-das} ) or a single control-tree for both types of threads ({\sc das}).
Additionally, for comparison purposes, %we add two additional lines: the ``ideal'' case line
%corresponding to the sum of the peak performances of the configurations 
%that use four cores of each of the two types in isolation; and 
we include the performance lines of the best 
%static {\sc sas} and 
{\sc ca-sas} strategy with a distribution ratio of 5.

\begin{figure}[th!]
\begin{center}
\begin{tabular}{c}
\begin{minipage}[c]{\textwidth}
\includegraphics[width=0.5\textwidth]{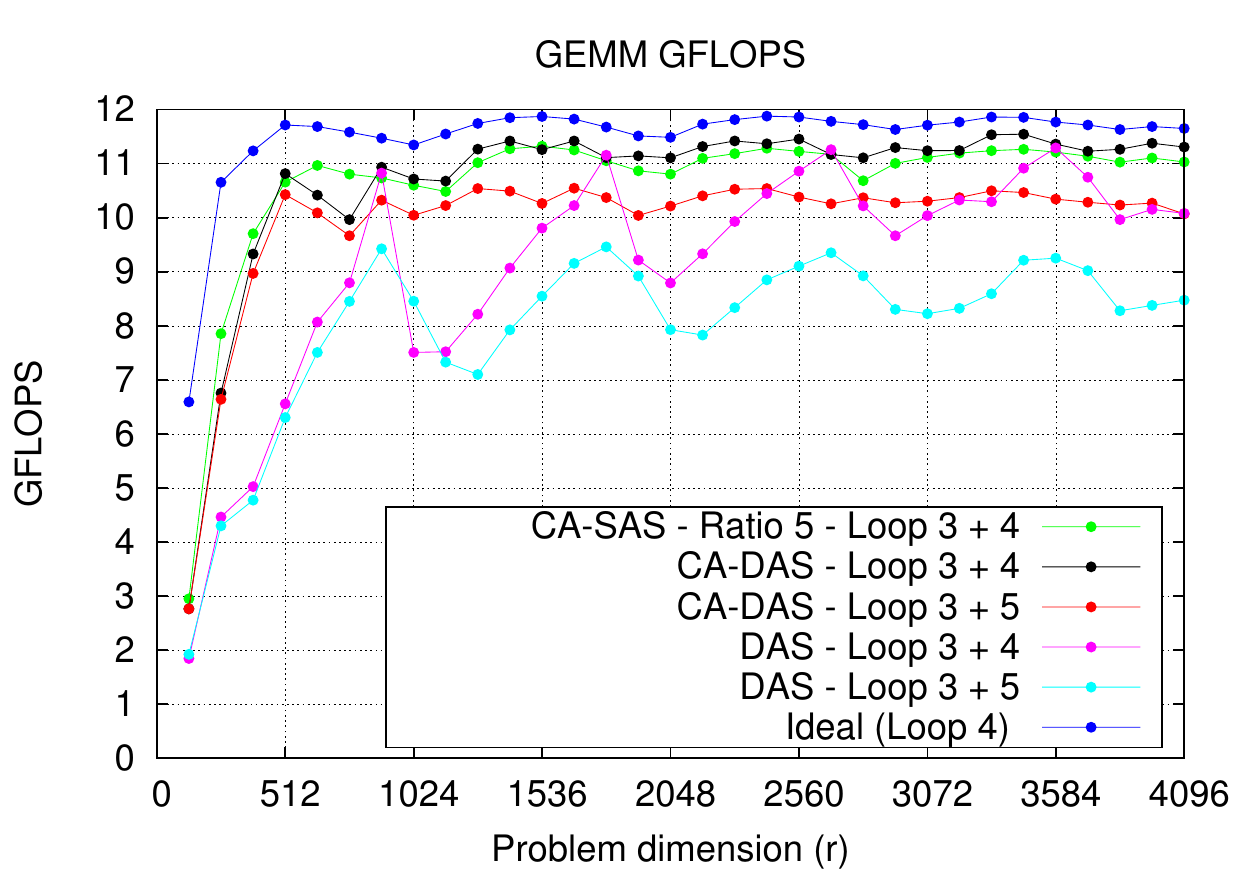}
\includegraphics[width=0.5\textwidth]{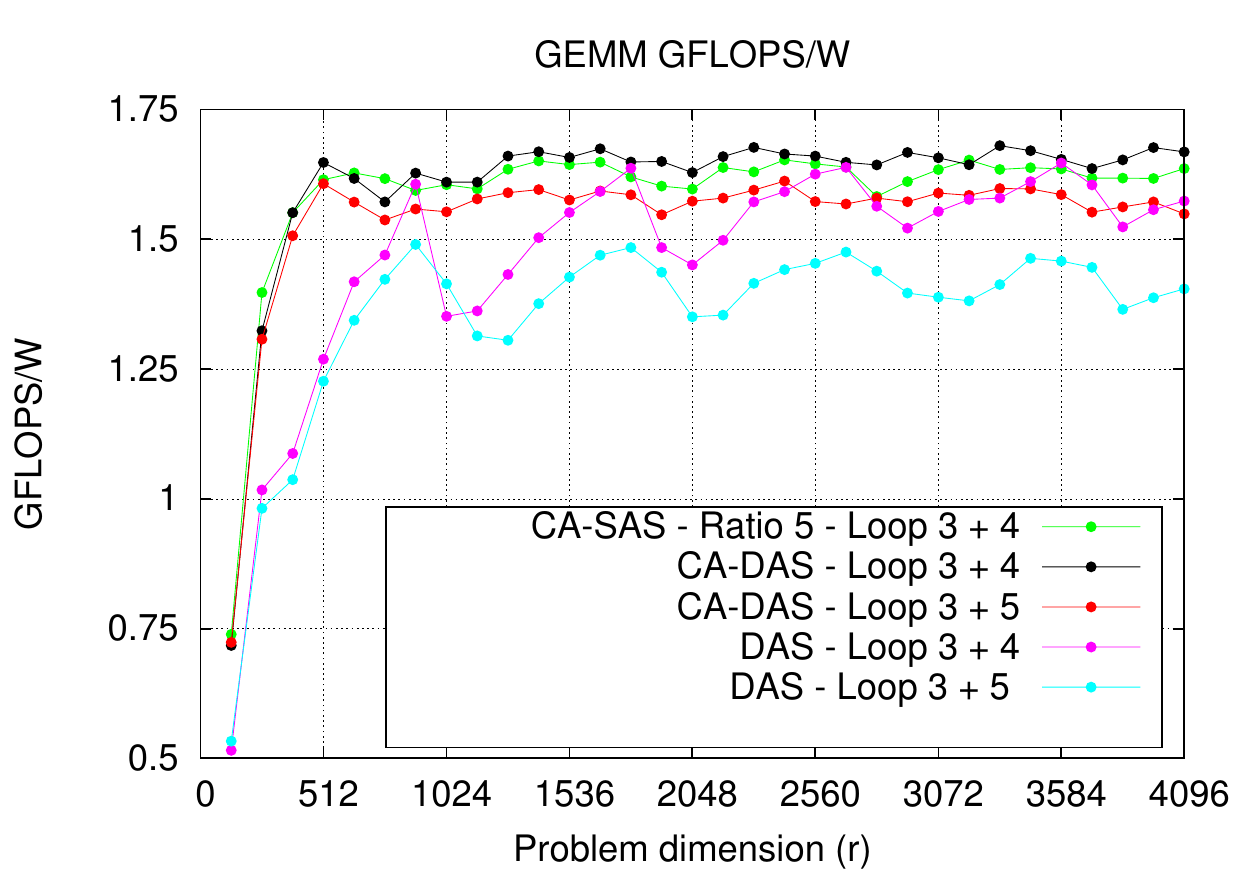}
\end{minipage}
\end{tabular}
\end{center}
\caption{Performance (left) and energy-efficiency (right) of the
	 {\sc ca-das} and {\sc das} versions of BLIS \gemm with a coarse-grain parallelization of
         Loop~3 and a fine-grain parallelization of either Loop~4 or Loop~5, using 4 threads per cluster in both cases.}
%\caption{%Performance (left) and energy-efficiency (right) of the BLIS DGEMM 
%         %distributing iterations of Loop~3 and four thread leveraged in Loop~5.}
%         Performance (left) and energy-efficiency (right) of the BLIS \gemm with static-asymmetric
%         scheduling and cache-aware configuration, from top to bottom parallelizing loops 1+4, 1+5, 3+4, and 3+5.}
%\label{fig:3-5-ASYM}
\label{fig:Dynamic}
\end{figure}

The plots in \figurename~\ref{fig:Dynamic} reveal that, for both metrics of interest, the best results are 
attained when the coarse-grain parallelization is dynamically applied to Loop~3 and the fine-grain parallelization 
is done at Loop~4. If the fine-grain parallelization is set across Loop~5, the results are inferior to those
reported for the static approach, since the amount of concurrency that can be extracted is lower for Loop~5 than
for Loop~4 (see \figurename~\ref{fig:Ratio5_ASYM} and the corresponding analysis for details).
On the other hand, the plots show that the use of two control-trees has a great impact on both metrics. 
The use of a common control-tree implies that the chunk size assigned to both types of threads is the same.
Therefore, due to the difference in performance of the Cortex-A7 and Cortex-A15 clusters, this produces a severe load 
unbalance for certain problem sizes.

\section{Conclusions}
\label{sec:conclusions}

We have proposed and evaluated several mechanisms to efficiently map the
framework for matrix multiplication integrated in the BLIS library to an asymmetric ARM big.LITTLE 
(Cortex A15+A7) SoC. Our results reveal excellent improvements in performance 
compared with a homogeneous implementation that operates exclusively on one type of core 
(either A15 or A7), and also with respect to multi-threaded implementations that
simply apply a symmetric workload distribution and do not take into account the different cache organization of the 
cores.

This is an important step towards a full BLAS implementation optimized for 
big.LITTLE architectures, which is a future goal in our research effort. 
While we believe that the approach applied to \gemm carries over
to the rest of the BLAS, there are a number of issues 
that need to be addressed to further increase performance and adaption to other (present
and future) asymmetric architectures. 
Among others, the most relevant factor is the adoption of different micro-kernels,
tuned to each type of core, in order to extract the maximum performance
for those asymmetric architectures. 
A port to a 64-bit ARMv8 architecture, and an experimental study on
architectures with different number of big/LITTLE cores are also key milestones in our roadmap.

\section*{Acknowledgments}

The researchers from Universitat Jaume~I 
were supported by project CICYT TIN2011-23283 of
MINECO and FEDER, the EU project FP7 318793 ``EXA2GREEN'' and the FPU
program of MECD.
The researcher from Universidad Complutense de Madrid 
was supported by project CICYT TIN2012-32180.


%  \bibliographystyle{elsarticle-num} 
%  \bibliography{enrique,energy,asymmetric}

%\section*{References}


%\begin{thebibliography}{00}
%%% Text of bibliographic item
%
%\bibitem{}
%
%\end{thebibliography}
\end{document}